\documentclass{article}

\usepackage[utf8]{inputenc}
\usepackage{amssymb}
\usepackage{mathtools}
\usepackage{bbm}
\usepackage[paper=a4paper,text={17cm,24cm},centering]{geometry}
\usepackage{graphicx}
\usepackage{enumitem}
\usepackage{hyperref}
\usepackage{subcaption}
\usepackage{cancel}
\usepackage{authblk}
\usepackage{booktabs}
\usepackage{placeins}
\usepackage{tikz}
\usetikzlibrary{arrows.meta,arrows}
\usepackage[compat=1.1.0]{tikz-feynman}
\tikzfeynmanset{/tikzfeynman/warn luatex=false}

\def\be{\begin{equation}}
\def\ee{\end{equation}}
\def\P{{\mathbbm P}}
\def\Psat{P_\text{sat}}
\def\Pc{P_0}

\usepackage{soul}

\title{Numerical study of Darcy's law of yield stress fluids\\
on a deep tree-like network}

\author[1]{Stéphane Munier}
\author[2]{Alberto Rosso}

\affil[1]{CPHT, CNRS, École polytechnique, Institut Polytechnique de Paris, 91120 Palaiseau, France}
\affil[2]{LPTMS, CNRS, Université Paris-Saclay, 91405 Orsay, France}

\date{\today}

\begin{document}

\maketitle

\begin{abstract}
Understanding the flow dynamics of yield stress fluids in porous media presents a substantial challenge. Both experiments and extensive numerical simulations frequently show a non-linear relationship between the flow rate and the pressure gradient, deviating from the traditional Darcy law. In this article, we consider a tree-like porous structure and utilize an exact mapping with the directed polymer (DP) with disordered bond energies on the Cayley tree. Specifically, we adapt an algorithm recently introduced by Brunet et al. [Europhys. Lett. {\bf 131}, 40002 (2020)] to simulate exactly the tip region of branching random walks with the help of a spinal decomposition, to accurately compute the flow on extensive trees with several thousand generations. Our results confirm the asymptotic predictions proposed by Schimmenti et al. [Phys. Rev. E {\bf 108}, L023102 (2023)], tested therein only for moderate trees of about 20 generations.
\end{abstract}


\section{Introduction}

The Darcy law describes the flow rate $Q$ of a Newtonian fluid through a cylinder of radius $R$ and length $L$, filled with a porous medium \cite{Darcy:1856}:
\begin{equation}\label{eqn:darcynewt}
Q = \frac{\kappa R^2}{\eta L}\Delta P\,.
\end{equation}
In this equation, $\Delta P$ is the pressure drop, namely the difference of the gauge pressures $P$ at the inlet and $P'\leq P$ at the the outlet of the cylinder, respectively. The parameter $\eta$ represents the fluid's viscosity, while $\kappa$, known as permeability, is dimensionally an area that characterizes the medium's ability to transmit fluid.\cite{Bear:1988,Sahimi:2011,Blunt:2017,Feder:2022}. In the absence of a porous medium, the flow rate is described by the famous Poiseuille law:
\begin{equation}\label{eq:Poiseuille-classical}
Q_{\text{Pois.}}(R) = \frac{\pi R^4}{8 \eta L}\Delta P\,.
\end{equation}
The permeability of a filled cylinder corresponds to an area much smaller than the  section, $\pi  R^2$. A simple model, proposed by H. Darcy, is based on the assumption that flow in a porous medium occurs through numerous thin, non-intersecting channels, each with a radius $R_{\rm ch} \ll R$ and of number $n^{\rm ch}$ per unit area. According to Poiseuille's law, the flow through the filled cylinder can be written as $Q = \pi R^2 n^{\rm ch} \times Q_{\text{Pois.}}(R_{\rm ch})$. This leads to an expression for permeability as $\kappa = (\pi^2/8) n^{\rm ch} R_{\rm ch}^4$, which can be rewritten, up to an unimportant numerical factor $1/8$, as the product of: (i) the fraction of unit area occupied by the fluid, $n^{\rm ch} \pi R_{\rm ch}^2$, a positive number smaller than one, and (ii) the area of the section of the microchannel $\pi R_{\rm ch}^2$. Although actual porous media feature a complex network of intersecting channels with varying shapes, the Darcy law holds true as long as the number of open channels $n^{\rm ch}$ remains independent of pressure.

This Darcy framework is significantly altered in the presence of yield stress fluids—fluids that behave like a solid below a given yield stress, $\sigma_Y$, and flow only when the stress exceeds this threshold. The consequence of yield stress is that the number of channels through which the fluid flows increases as the pressure drop $\Delta P$ applied at the ends of the throat increases. Three regimes are observed in experiments and numerical simulations:
(i) no flow is observed below a critical pressure drop $\Pc$;
(ii)  a non linear flow is observed when the pressure drop is larger than $\Pc$  \cite{Roux:1987,Liu:2019}, the non linearity being related to the growing number of open channels above threshold;
(iii) only above a saturation pressure drop $\Psat \gg \Pc$, does the flow revert to linear growth,  the linear growth with the standard permeability~$\kappa$~\cite{Barnes:2000}.

Pore-network models are the simplest models that capture these three regimes. They feature a graph structure in which the nodes represent large open pores with well-defined pressure, and the edges are throats  connecting the pores. Establishing how the Poiseuille law is modified in the presence of a yield stress remains an open challenge. In this study, we examined the simplest modification to the Poiseuille law (\ref{eq:Poiseuille-classical}):
\be
Q= \frac{\pi R^4}{8\eta L}\times\left(\Delta P-\tau\right)_+\,,
\label{eq:elementary-flow-0}
\ee
where $(x)_+\equiv \max(0,x)$ and  the  threshold pressure drop is $\tau=L \sigma_Y /R$  \cite{Poiseuille:1840}.
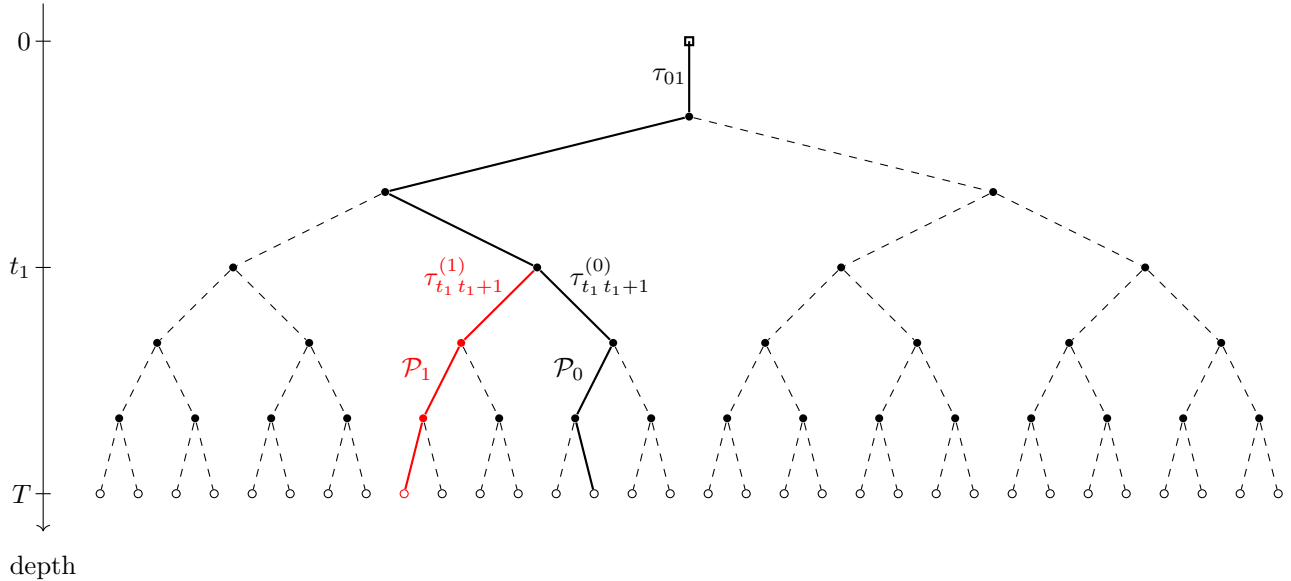
\begin{figure}
    \begin{center}
    \begin{tikzpicture}[
  level/.style={sibling distance=8cm/(2^(#1-1))},level distance=1cm,
  every node/.style={fill=black,circle,inner sep=0pt,minimum size=3pt},
  every child/.style={dashed},
  ]
  \node (root) {}
  child [white]{ node (child2) {}
    child [black]{ node {}
      child { node {}
        child { node {}
          child { node[draw,solid,fill=none] {} }
          child { node[draw,solid,fill=none] {} }
        }
        child { node {}
          child { node[draw,solid,fill=none] {} }
          child { node[draw,solid,fill=none] {} }
        }
      }
      child { node {}
        child { node {}
          child { node[draw,solid,fill=none] {} }
          child { node[draw,solid,fill=none] {} }
        }
        child { node {}
          child { node[draw,solid,fill=none] {} }
          child { node[draw,solid,fill=none] {} }
        }
      }
    }
    child { node (child3) {}
      child { node[fill=red] (child14) {}
        child { node[fill=red] (child15) {}
          child { node[draw,solid,red,fill=none] (child16) {} }
          child [black] { node[draw,solid,fill=none] {} }
        }
        child [black] { node {}
          child { node[draw,solid,fill=none] {} }
          child { node[draw,solid,fill=none] {} }
        }
      }
      child { node (child4) {}
        child { node (child5) {}
          child [black] { node[draw,solid,fill=none] {} }
          child [black] { node[draw,solid,fill=none] (child6) {} }
        }
        child [black] { node {}
          child { node[draw,solid,fill=none] {} }
          child { node[draw,solid,fill=none] {} }
        }
      }
    }
  }
  child { node {}
    child { node {}
      child { node {}
        child { node {}
          child { node[draw,solid,fill=none] {} }
          child { node[draw,solid,fill=none] {} }
        }
        child { node {}
          child { node[draw,solid,fill=none] {} }
          child { node[draw,solid,fill=none] {} }
        }
      }
      child { node {}
        child { node {}
          child { node[draw,solid,fill=none] {} }
          child { node[draw,solid,fill=none] {} }
        }
        child { node {}
          child { node[draw,solid,fill=none] {} }
          child { node[draw,solid,fill=none] {} }
        }
      }
    }
    child { node {}
      child { node {}
        child { node {}
          child { node[draw,solid,fill=none] {} }
          child { node[draw,solid,fill=none] {} }
        }
        child { node {}
          child { node[draw,solid,fill=none] {} }
          child { node[draw,solid,fill=none] {} }
        }
      }
      child { node {}
        child { node {}
          child { node[draw,solid,fill=none] {} }
          child { node[draw,solid,fill=none] {} }
        }
        child { node {}
          child { node[draw,solid,fill=none] {} }
          child { node[draw,solid,fill=none] {} }
        }
      }
    }
  };
  \coordinate (root0) at (0,1cm);
  \draw[fill=none,black,thick] (root0) node[draw,rectangle,fill=none]{} -- node[draw=none,fill=none,anchor=east]{$\tau_{01}$} (root);
  \draw[fill=none,solid,black,thick] (root)--(child2)--(child3)--node[fill=none,anchor=220]{$\tau_{t_1\,t_1+1}^{(0)}$}(child4)--node[fill=none,anchor=-30]{${\cal P}_0\ $}(child5)--(child6);
  \draw[fill=none,solid,red,thick] (child3)--node[fill=none,anchor=-40,red]{$\tau_{t_1\,t_1+1}^{(1)}$}(child14)--node[fill=none,anchor=-30]{${\cal P}_1\ $}(child15)--(child16);

  \draw[->] (-8.5,1.5) -- (-8.5,-5.5) node[anchor=north,draw=none,fill=none] {depth};
  \draw (-8.6,1) node[fill=none,anchor=east] {$0$} -- (-8.4,1);
    \draw (-8.6,-2) node[fill=none,anchor=east] {$t_1$} -- (-8.4,-2);
  \draw (-8.6,-5) node[fill=none,anchor=east] {$T$} -- (-8.4,-5);

\end{tikzpicture}
    \end{center}
    \caption{Tree-like network of hydraulic pores and throats (or of directed polymer configurations) of height $T=6$. The edges represent throats (or bonds between monomers), each characterized by a random threshold pressure differential (or energy) $\tau$ [see Eq.~(\ref{eq:elementary-flow})]. To illustrate our discussion, we have highlighted two channels (or configurations), ${\cal P}_0$ and ${\cal P}_1$ of overlap $t_1$ (shown with solid lines).  In a particular disorder realization, these are the first channels to open as increasing pressure is applied to the inlet, marked by a square.}
    \label{fig:Cayley-tree}
\end{figure}
We shall assume the overall coefficient fixed, and randomly draw the thresholds $\tau$ as independent and identically distributed variables.

Following Refs.~\cite{Schimmenti:2023,Schimmenti:2023_supplemental}, the hydraulic network is organized as a single inlet throat feeding a perfect binary tree with a total height of $T-1$ (see Fig.~\ref{fig:Cayley-tree}). The set of the $2^{T}-1$ variables $\tau$ that specify each sample can be interpreted as ``frozen disorder'' in statistical physics terms.

The tree-like network of pipes maps exactly onto the configuration space of directed polymers (DP) \cite{Mezard:1986,Derrida:1988} in the infinite-dimensional limit. The elementary throats correspond to bonds between monomers, and the $\tau$ variables  represent their energies in that context. Therefore, as noted in Refs.~\cite{Liu:2019,Schimmenti:2023}, many tools developed for the DP problem can be adapted to the Darcy problem. Understanding the former has benefited from the observation that configuration spaces are generated by branching random walks in the universality class of branching Brownian motion \cite{Derrida:1988}. The main goal of this paper is to demonstrate how an algorithm developed in the latter context~\cite{Brunet:2020yiq} can be applied to this fluid-mechanics problem to numerically investigate it on very large tree-like networks. Specifically, this algorithm will allow us to test a prediction proposed in Ref.~\cite{Schimmenti:2023}, which led to asymptotic analytical expressions for a set of relevant observables.



The paper is organized as follows. Section~\ref{sec:fluid-mechanics} provides an alternative and pedagogical derivation of the algorithm used in Ref.~\cite{Schimmenti:2023}. This algorithm computes the flow curve for moderate trees $T\sim 20$ because it requires the prior generation of the full network ($2^{T-1}$ branches). In Section~\ref{sec:advanced-algo}, we introduce a new algorithm to compute the flow curve, inspired by the one described in Ref.~\cite{Brunet:2020yiq}. Our new algorithm can compute the flow curve for trees with $T\sim 10^3$ because it draws only the sequence of the first open channels, without prior generation of the full network. Our results are presented and discussed in Section~\ref{sec:results}, and our conclusions are drawn in Section~\ref{sec:conclusion}. Two appendices provide technical details.


\section{Basic algorithm to compute the flow}
\label{sec:fluid-mechanics}

According to the discussion in the Introduction, we  model the pressure dependence of the flow in each elementary throat using  Eq.~(\ref{eq:elementary-flow-0}). By an appropriate choice of units, we can set the overall constant to unity, to get
\be
Q=\left(P-P'-\tau\right)_+\,,
\label{eq:elementary-flow}
\ee
where we have explicitly defined the pressure drop, $\Delta P$ in Eq.~(\ref{eq:elementary-flow-0}), as the difference between the inlet gauge pressure, $P$, and the outlet gauge pressure, $P'$. Furthermore, the Kirchhoff's law \cite{Kirchhoff:1868} is assumed at each node of the tree that connects two outgoing throats to an incoming one. Specifically, the sum of the outgoing flows $Q^{(0)}$ and $Q^{(1)}$ is equal to the incoming flow $Q$:
\be
Q=Q^{(0)}+Q^{(1)}.
\label{eq:Kirchhoff}
\ee
Equations~(\ref{eq:elementary-flow}) and (\ref{eq:Kirchhoff}) are sufficient to determine the properties of the flow for any disorder realization.

The goal of this section is to design the simplest algorithm to compute the flow curve $Q_T(P)$  as a function of the pressure $P$ applied at the inlet, for a given realization of the disorder in a network tree of height $T$. 


\subsection{Warm-up: the simple trees with \texorpdfstring{$T=1$}{T=1} and \texorpdfstring{$T=2$}{T=2}}

We begin by presenting detailed analytical calculations of the flow in elementary cases with $T=1$ and $T=2$, formulated in a way that facilitates understanding the more general case.

\begin{figure}[h]
\begin{center}
    \begin{subfigure}[t]{0.3\textwidth}
        \begin{center}
        \begin{tikzpicture}      
            \draw[black,thick] (0,1.5) node[draw,rectangle,inner sep=0pt,minimum size=5pt,fill=none]{} -- (0,0) node[draw,circle,inner sep=0pt,minimum size=5pt,fill=none]{};
            \node[draw,anchor=east] at (-0.3,1.5) {$P$};
            \node[draw,anchor=east] at (-0.3,0) {$P'=0$};
            \node[anchor=west] at (0.05,0.7) {$Q_1(P)$};
            \node[anchor=east] at (-0.05,0.7) {$\tau_{01}$};
            \draw[white] (0,0) -- (0,-1.5);
        \end{tikzpicture}
        \end{center}
        \caption{Height $T=1$.}
        \label{fig:tree-T=1}
    \end{subfigure}
    \hfil
    \begin{subfigure}[t]{0.3\textwidth}
    \begin{center}
    \begin{tikzpicture}
    \draw[black,thick] (0,1.5) node[draw,rectangle,inner sep=0pt,minimum size=5pt,fill=none]{} -- (0,0);
    \draw[black,thick] (0,0) -- (-1.5,-1.5) node[draw,circle,inner sep=0pt,minimum size=5pt,fill=none]{};
    \draw[black,thick] (0,0) -- (1.5,-1.5) node[draw,circle,inner sep=0pt,minimum size=5pt,fill=none]{};
    \node[draw,anchor=east] at (-0.3,1.5) {$P$};
    \node[anchor=east] at (-0.05,0.7) {$\tau_{01}$};
    \node[anchor=west] at (0.05,0.7) {$Q_2(P)$};
    \node[anchor=west] at (1,-0.8) {$\tau_{12}^{(1)}$};
    \node[anchor=east] at (-1,-0.8) {$\tau_{12}^{(0)}$};
    \node[draw,anchor=east] at (-0.3,0.1) {$P'$};
    \filldraw[black] (0,0) circle (2pt);
    \end{tikzpicture}
    \end{center}
    \caption{Height $T=2$.}
    \label{fig:tree-T=2}
    \end{subfigure}
    \end{center}
    \caption{Tree-like networks of smallest heights.}
\end{figure}
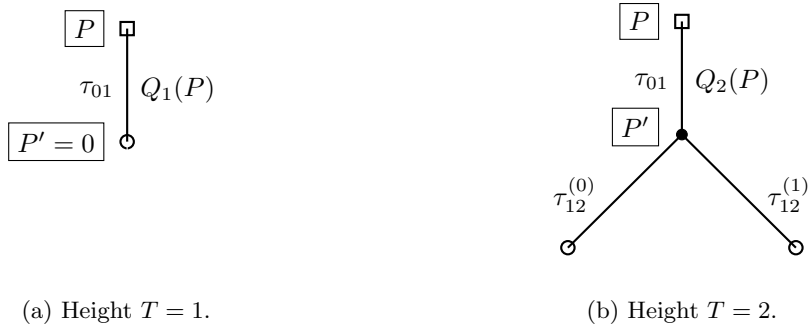

In the  case $T=1$, there is only one throat, characterized by the threshold $\tau_{01}$; see Fig.~\ref{fig:tree-T=1}. The flow as a function of the pressure is then just given by Eq.~(\ref{eq:elementary-flow}), in which $P'$ is the pressure at the free outlet of the network, namely $P'=0$:
\be
Q_1(P)=\begin{dcases}
0 & \text{for $P\leq\tau_{01}$},\\
P-\tau_{01} & \text{for $P>\tau_{01}$}.
\end{dcases}
\quad
\begin{tikzpicture}[baseline=0.4cm]
\draw[->] (0,0)--(0,1) node [anchor=south]{\small $Q_1(P)$};
\draw[->] (0,0)--(2,0) node [anchor=west]{\small $P$};
\draw[thick] (0,0)--(1,0)--(2,1);
\draw (1,-0.1)node[anchor=north]{\small $P_0=\tau_{01}$}--(1,0.1);
\end{tikzpicture}
\ee
Throughout, we denote $P_0$ as the disorder-dependent minimum pressure that needs to be applied at the inlet of a network to establish the flow. In the present case, obviously, $P_0=\tau_{01}$.

In the case $T=2$, the network has one node connecting three throats; see Fig.~\ref{fig:tree-T=2}. 
Let us call $P'$ the pressure at the node, and call $\tau_{01}$, $\tau_{12}^{(0)}$ and $\tau_{12}^{(1)}$ the thresholds that characterize the incoming and the two outgoing throats respectively. For the fluid to flow, the two following conditions must be satisfied: 
\be
P-P'>\tau_{01} \quad \text{and} \quad P'>\min\left(\tau_{12}^{(0)},\tau_{12}^{(1)}\right).
\ee
This implies $P>P_0$, where
\be
P_0=\tau_{01}+\min\left(\tau_{12}^{(0)},\tau_{12}^{(1)}\right).
\label{eq:P0_T=2}
\ee
In words, Eq.~(\ref{eq:P0_T=2}) says that $P_0$ minimizes the sum of the thresholds $\tau$ along the directed paths connecting the root to the leaves of the tree. This definition of $P_0$ actually generalizes to trees of any height.

Let us set the pressure $P$ at the inlet just above the threshold $P_0$. Assuming for the time being that $\tau_{12}^{(0)}\neq \tau_{12}^{(1)}$ and hence that $P_0$ is not degenerate, the flow reads
\be
Q_2(P)=P-P'-\tau_{01}=P'-\min\left(\tau_{12}^{(0)},\tau_{12}^{(1)}\right).
\ee
The first equality is just Eq.~(\ref{eq:elementary-flow}) applied to the incoming throat. The second one follows from the conservation of the flow at the node Eq.~(\ref{eq:Kirchhoff}), in the case in which only one outgoing channel is open, and from Eq.~(\ref{eq:elementary-flow}) applied to the open outgoing throat. We may solve for $P'$, arriving at the following determination:
\be
P'=\frac12\left[P-\tau_{01}+\min\left(\tau_{12}^{(0)},\tau_{12}^{(1)}\right)\right].
\ee
It is then straightforward to deduce $Q_2(P)$:
\be
Q_2(P)=\frac12\left[P-\tau_{01}-\min\left(\tau_{12}^{(0)},\tau_{12}^{(1)}\right)\right]_+=\frac12\left(P-P_0\right)_+.
\ee
This equation holds for pressures below the threshold $P_1$ for the opening of the second channel. 

The flow above $P_1$ is determined using again Eq.~(\ref{eq:elementary-flow}) and the Kirchhoff law~(\ref{eq:Kirchhoff}) at the node, now assuming that both channels are open:
\be
Q_2(P)=P-P'-\tau_{01}=2P'-\tau_{12}^{(0)}-\tau_{12}^{(1)}.
\label{eq:Q2}
\ee
Solving for $P'$, 
\be
P'=\frac13\left(P+\tau_{12}^{(0)}+\tau_{12}^{(1)}-\tau_{01}\right).
\label{eq:Pp_T=2}
\ee
We evaluate the threshold pressure $P_1$ based on the requirement that, for the flow to establish in both branches of the tree, the pressure $P'$ at the node must exceed the threshold pressures of both outgoing throats:
\be
P'\geq\max\left(\tau_{12}^{(0)},\tau_{12}^{(1)}\right).
\ee
The pressure $P_1$ coincides with the value of the pressure $P$ at the inlet such that this inequality for $P'$ is saturated. Equation~(\ref{eq:Pp_T=2}) then yields
\be
P_1=P_0+2\left[\max\left(\tau_{12}^{(0)},\tau_{12}^{(1)}\right)
-\min\left(\tau_{12}^{(0)},\tau_{12}^{(1)}\right)\right].
\label{eq:P1_T=2}
\ee
Now, inserting the expression for $P'$ in Eq.~(\ref{eq:Pp_T=2}) into one of the equations~(\ref{eq:Q2}) for $Q_2(P)$, we get the expression of the flow:
\be
Q_2(P)=\frac{2}{3}\left[P-\tau_{01}-\frac12\left(\tau_{12}^{(0)}+\tau_{12}^{(1)}\right)\right].
\ee
All in all, we have found that the flow depends on the applied pressure as
\be
Q_2(P)=\begin{dcases}
0 & \text{for $P\leq P_0$}\\
\frac12\left[P-\tau_{01}-\min\left(\tau_{12}^{(0)},\tau_{12}^{(1)}\right)\right] & \text{for $P_0<P\leq P_1$}\\
\frac{2}{3}\left[P-\tau_{01}-\frac12\left(\tau_{12}^{(0)}+\tau_{12}^{(1)}\right)\right]
& \text{for $P>P_1$},
\end{dcases}
\quad
\begin{tikzpicture}[baseline=0.6cm]
\draw[->] (0,0)--(0,1.5) node [anchor=south]{\small $Q_2(P)$};
\draw[->] (0,0)--(2.5,0) node [anchor=west]{\small $P$};
\draw[thick] (0,0)--(0.5,0)--(1.5,0.5)--(2.5,0.5+1*2/3);
\draw[dashed] (1.5,0.5)--(2.5,1);
\draw (0.5,-0.1)node[anchor=north]{\small $P_0$}--(0.5,0.1);
\draw (1.5,-0.1)node[anchor=north]{\small $P_1$}--(1.5,0.1);
\draw[dotted](1.5,0.1)--(1.5,0.5);
\end{tikzpicture}
\label{eq:Q_2(P)}
\ee
where the thresholds $P_0$ and $P_1$ are given in Eqs.~(\ref{eq:P0_T=2}) and~(\ref{eq:P1_T=2}) respectively. Let us comment that the degenerate case $\tau_{12}^{(0)}=\tau_{12}^{(1)}$ does not require a special treatment: $P_0=P_1=\tau_{01}+\tau_{12}^{(0)}$, and the formula we have just established applies, the second distinguished case being simply no longer necessary as the pressure interval in which it applies becomes trivial.

We see that $Q_2(P)$ is a piece-wise linear and continuous function. It is not difficult to figure out that this property generalizes for trees of arbitrary heights.


\subsection{Trees of arbitrary heights}

Given a network of size $T$ with all the $2^T-1$ elementary thresholds $\tau$, we take the following generic Ansatz for the flow at the inlet of the complete tree, or of sub-trees of it, as that of an effective Darcy law:
\be
Q(P)=\kappa_{\text{eff}}(P)\left[P-P^*_{\text{eff}}(P)\right].
\label{eq:Q(P)}
\ee
We shall call the proportionality factor $\kappa_\text{eff}(P)$ as the {\em effective permeability} of the considered (sub-)network, and $P^*_{\text{eff}}(P)$ as the {\em effective offset}. These two flow parameters   are non-decreasing piece-wise constant functions of $P$.  Their  discontinuities occur at the pressures  at which new channels  open. Our goal is to compute recursively $\kappa_{\text{eff}}(P)$ and $P_{\text{eff}}^*(P)$ for a pressure $P$ such that there are, say, $n^\text{ch}$ open channels. 


\subsubsection{Flow parameters at a given node}

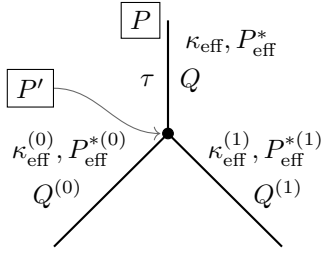
\begin{figure}
    \begin{center}
    \begin{tikzpicture}
    \draw[black,thick] (0,1.5) -- (0,0);
    \draw[black,thick] (0,0) -- (-1.5,-1.5);
    \draw[black,thick] (0,0) -- (1.5,-1.5);
    \node[draw,anchor=east] at (-0.1,1.5) {$P$};
    \node[anchor=west] at (0.1,1.2) {$\kappa_\text{eff},P_\text{eff}^*$};
    \node[anchor=west] at (0.03,0.7) {$Q$};
    \node[anchor=east] at (-0.05,0.7) {$\tau$};
    \node[anchor=west] at (0.4,-0.2) {$\kappa_\text{eff}^{(1)},P_\text{eff}^{*(1)}$};
    \node[anchor=west] at (1,-0.8) {$Q^{(1)}$};
    \node[anchor=east] at (-0.4,-0.2) {$\kappa_\text{eff}^{(0)},P_\text{eff}^{*(0)}$};
    \node[anchor=east] at (-1,-0.8) {$Q^{(0)}$};
    \node[draw,anchor=east] at (-1.5,0.6) {$P'$};
    \draw[->,black!60,thin,out=0,in=180] (-1.5,0.6) to (-0.1,0.);
    \filldraw[black] (0,0) circle (2pt);
    \end{tikzpicture}
    \end{center}
    \caption{A typical node in the network. $P$ denotes the pressure applied at the inlet of the ingoing throat, $P'$ is the pressure at the inlet of the two outgoing throats. $Q$, $Q^{(0)}$ and $Q^{(1)}$ are the flows in these respective throats. They depend on the pressures $P$, $P'$ according to the Darcy law~(\ref{eq:Q(P)}), with the indicated parameters, and are related by the Kirchhoff law~(\ref{eq:Kirchhoff}).}
    \label{fig:typical-node}
\end{figure}

Let us start by assuming that the sub-tree of all open channels is known, and consider a generic node in the latter. We label the throats it connects as in Fig.~\ref{fig:typical-node}, and write the Kirchhoff law~(\ref{eq:Kirchhoff}) at this node.
We then replace the flows in the right-hand side of the latter by Eq.~(\ref{eq:Q(P)}), with $P$ set to the pressure $P'$ at the node and with appropriate superscripts for the parameters. As for $Q$ in the left-hand side of Eq.~(\ref{eq:Kirchhoff}), we shall eventually replace it by Eq.~(\ref{eq:elementary-flow}). 

We first address the case in which only one of the outgoing channels is open, say the one labeled $(0)$:
\be
Q=P-P'-\tau=\kappa_{\text{eff}}^{(0)}\left(P'-P^{*(0)}_{\text{eff}}\right).
\label{eq:Q-generic-node}
\ee
$Q$ in the left-hand side will eventually depend on $P$ only. The effective permeabilities and offsets in the right-hand side depend on $P'$ as piece-wise constant functions. Hence in the restricted range of $P$ (and in the corresponding one of $P'$) of interest, the second and third members form a linear equation for $P'$. Solving the latter, inserting the solution back into one of the alternative expressions for $Q$, we get for the flow
\be
Q=\frac{\kappa_\text{eff}^{(0)}}{1+\kappa_\text{eff}^{(0)}}\left(P-\tau-P_\text{eff}^{*(0)}\right).
\ee
Recalling that $Q(P)$ itself obeys Darcy's law~(\ref{eq:Q(P)}), it is straightforward to identify the parameters that characterize the flow at the inlet:
\be
\kappa_\text{eff}^{(0\wedge\cancel{1})}=\frac{\kappa_\text{eff}^{(0)}}{1+\kappa_\text{eff}^{(0)}}
\quad\quad
P_\text{eff}^{*{(0\wedge\cancel{1})}}=\tau+P_\text{eff}^{*(0)}\,.
\label{eq:flow-parameters-1}
\ee
We introduced self-explanatory superscripts. It is of course enough to exchange the superscripts $(0)$ and $(1)$ to get the case in which sub-channel $(1)$ is open and sub-channel $(0)$ is closed

Next, let us assume that $P$ is sufficiently large, such that the two outgoing channels are open. Instead of Eq.~(\ref{eq:Q-generic-node}), we now have the following equation:
\be
Q=P-P'-\tau=\kappa_{\text{eff}}^{(0)}\left(P'-P^{*(0)}_{\text{eff}}\right)+\kappa_{\text{eff}}^{(1)}\left(P'-P^{*(1)}_{\text{eff}}\right).
\ee
Solving for $P'$ and inserting its solution back into the expression for $Q$, we arrive at
\be
Q=\frac{\kappa_\text{eff}^{(0)}+\kappa_\text{eff}^{(1)}}{1+\kappa_\text{eff}^{(0)}+\kappa_\text{eff}^{(1)}}\left(P-\tau-\frac{\kappa_\text{eff}^{(0)}P_\text{eff}^{*(0)}+\kappa_\text{eff}^{(1)}P_\text{eff}^{*(1)}}{\kappa_\text{eff}^{(0)}+\kappa_\text{eff}^{(1)}}\right).
\ee
Comparing to Eq.~(\ref{eq:Q(P)}), we easily obtain the following relations between the effective flow parameters:
\be
\kappa_\text{eff}^{(0\wedge1)}=\frac{{\kappa_\text{eff}^{(0)}+\kappa_\text{eff}^{(1)}}}{1+{\kappa_\text{eff}^{(0)}+\kappa_\text{eff}^{(1)}}}
\quad\quad
P_\text{eff}^{*(0\wedge1)}=\tau+\frac{\kappa_\text{eff}^{(0)}P_\text{eff}^{*(0)}+\kappa_\text{eff}^{(1)}P_\text{eff}^{*(1)}}{\kappa_\text{eff}^{(0)}+\kappa_\text{eff}^{(1)}}.
\label{eq:flow-parameters-2}
\ee
We take the convention that the effective permeability is null for closed channels. We easily check that in the case in which only channel $(0)$ is open, Eq.~(\ref{eq:flow-parameters-2}) boils down to Eq.~(\ref{eq:flow-parameters-1}). Furthermore, the open throats located at the maximal depth, i.e. at the leaves of the sub-tree of open channels, have effective permeability $\kappa_\text{eff}=1$, and effective threshold $P_\text{eff}^*$ equal to the elementary threshold of the considered throat. Then, iterating Eqs.~(\ref{eq:flow-parameters-1}) and~(\ref{eq:flow-parameters-2}) upward from the leaves of the open sub-tree to its root enables one to determine the parameters $\kappa_{\text{eff}}$ and $P_{\text{eff}}^{*}$ of all sub-channels, and, eventually, of the complete tree.


\subsubsection{Threshold pressures} 

We recall that the expressions of the flow parameters just established are valid for pressures in the limited range in which the sub-tree we have singled out is that of all open channels. Let us now relate the threshold pressures of sub-trees rooted in the different throats joining at a given node. 

We call $P^{(0)}_\text{thr}$ and $P^{(1)}_\text{thr}$ the minimum pressures to apply at the inlet of the outgoing throats, for the sub-trees they root to be independently open. We compute the corresponding minimum pressure $P^{(0\wedge1)}_\text{thr}$ required at the inlet of the incoming throat to have a flow in both sub-trees. As above, we write the incoming flow for that pressure in two different ways:
\be
Q=P^{(0\wedge1)}_\text{thr}-\max\left(P^{(0)}_\text{thr},P^{(1)}_\text{thr}\right)-\tau
=\left(\kappa^{(0)}_\text{eff}+\kappa_\text{eff}^{(1)}\right)\max\left(P^{(0)}_\text{thr},P^{(1)}_\text{thr}\right)-\left(\kappa^{(0)}_\text{eff}P_\text{eff}^{*(0)}+\kappa^{(1)}_\text{eff}P_\text{eff}^{*(1)}\right)\,.
\ee
From the equality of the second and third members in this equation, we deduce the value of the threshold pressure:
\be
P^{(0\wedge1)}_\text{thr}=\tau+\left(1+\kappa_\text{eff}^{(0)}+\kappa_\text{eff}^{(1)}\right){\max\left(P^{(0)}_\text{thr},P^{(1)}_\text{thr}\right)}-\left(\kappa_\text{eff}^{(0)}P_\text{eff}^{*(0)}+\kappa_\text{eff}^{(1)}P_\text{eff}^{*(1)}\right).
\label{eq:P0and1}
\ee

The case in which only one outgoing channel, say $(0)$, is open is obtained by substituting $\max\left(P^{(0)}_\text{thr},P^{(1)}_\text{thr}\right)$ with $P^{(0)}_\text{thr}$, and setting $\kappa_\text{eff}^{(1)}$ to zero in this equation. The corresponding threshold pressure, that we shall denote by $P^{(0\wedge\cancel{1})}_\text{thr}$, reads
\be
P^{(0\wedge\cancel{1})}_\text{thr}=\tau+P^{(0)}_\text{thr}+
\kappa_\text{eff}^{(0)}\left(P^{(0)}_\text{thr}-P_\text{eff}^{*(0)}\right).
\label{eq:P0andnot1}
\ee
Obviously, the threshold pressure $P^{(\cancel{0}\wedge{1})}_\text{thr}$, corresponding to the case in which $(0)$ is closed and $(1)$ is open, is obtained by interchanging the superscripts $(0)$ and $(1)$ in this formula.

It will prove useful to have in mind the particular case of one-branch sub-trees, for which the threshold pressures $P_\text{thr}$ coincide with the effective pressures $P^*_\text{eff}$. According to Eq.~(\ref{eq:flow-parameters-1}), the latter just amount to the sums of the elementary thresholds $\tau$ along the branch of the considered sub-trees.

We iterate Eqs.~(\ref{eq:P0and1}) and~(\ref{eq:P0andnot1}) from the leaves to the root to determine the minimum pressure that needs to be exerted at the inlet of the network to open all channels of that sub-tree.


\subsubsection{One- and two-branch trees}

Here, we shall work out the expression for the flow parameters and of the threshold pressures in the cases $n^\text{ch}=1$ and $2$, but now for a tree of arbitrary height $T$. 

Considering a generic directed path ${\cal P}$ that connects the root of the tree to some leaf, and its segment ${\cal P}_{[t_1,t_2]}$ between the depths $t_1$ and $t_2$, we shall introduce the convenient notation
\be
\varepsilon^{({\cal P})}_{t_1\,t_2}\equiv \sum_{I\in{{\cal P}_{[t_1,t_2]}}} \tau_I
\label{eq:def-varepsilon-path}
\ee
for the path sum of the elementary thresholds along this segment. The indices $I$ over which the sum runs label the elementary segments the path is constructed of. We will also be led to use a notation such as 
\be
\min_{{\cal P}_{[t,T]}}\varepsilon_{0\,T}^{({\cal P})}\ ,
\ee
by which we mean the minimum of the quantity $\varepsilon_{0\,T}^{({\cal P})}$, varying the segment of the directed path ${\cal P}$ between the depths $t$ and $T$ (i.e. the nodes between $t+1$ and $T$) in all possible ways on the given tree, while keeping fixed the segment and all its nodes between the depths $0$ and $t$.

Let us address the single-branch case $n^\text{ch}=1$, consisting of the directed path ${\cal P}_0$. We may easily compute the flow parameters for sub-channels rooted at any node along the branch ${\cal P}_0$. Iterating Eqs.~(\ref{eq:flow-parameters-1}), we get
\be
\kappa_{\text{eff}\,t}=\frac{1}{T-t},\quad P^*_{\text{eff}\,t}=\varepsilon^{({\cal P}_0)}_{t\,T}
\label{eq:keff-Peff-nch=1}
\ee
for the sub-channel ${\cal P}_{0\,[t,T]}$ rooted at the node located at depth $t$ on ${\cal P}_0$, where $t$ is such that $0\leq t<T$. The corresponding threshold pressures $P_{\text{thr}\,t}$ identify to $P^*_{\text{eff}\,t}$, as observed in the previous section. 

If we want to require ${\cal P}_0$ to be the first path (or one of the first paths, in case of degeneracies) that opens when the pressure at the network inlet is increased from zero, it must obviously be a path that minimizes $P_{\text{eff}\,t=0}^*=\varepsilon^{({\cal P}_0)}_{0\,T}$. We denote the minimum effective threshold by $P_0$. If there is only one path that realizes the minimum, which is always the case if the distribution $p(\tau)$ of the $\tau$'s is continuous, then the flow as a function of the pressure $P$ at the inlet follows from Darcy's law~(\ref{eq:Q(P)}), with the parameters read off the previous equation after having set $t=0$:
\be
Q_T(P)=\frac{1}{T}\left(P-P_0\right)_+.
\label{eq:Q_T(P)-1-branch}
\ee
This relation holds only below the pressure above which the next channel opens.

We now turn to the case $n^\text{ch}=2$. We shall assume that the second branch is attached at a node at depth~$t_1$, and we denote by ${\cal P}_{1\hat t_1}$ the directed path from the root to the leaves of the tree that includes this branch. This path has obviously an overlap of size $t_1$ with ${\cal P}_0$, which amounts to all throats between the inlet of the network and the node at depth $t_1$. 

The flow parameters of the sub-trees rooted in nodes of depth $t\geq t_1$ are just those of the single-branch case for each of the channels. If $t<t_1$ instead, we apply Eq.~(\ref{eq:flow-parameters-2}), and then Eq.~(\ref{eq:flow-parameters-1}), possibly multiply. We find
\be
\kappa_{\text{eff}\,t}=\frac{2}{{T+t_1}-2t},\quad P_{\text{eff}\,t}^*=\varepsilon^{({\cal P}_0)}_{t\,t_1}+\frac12\left(\varepsilon^{({\cal P}_0)}_{t_1\,T}+\varepsilon^{({\cal P}_{1\hat t_1})}_{t_1\,T}\right).
\label{eq:kappa_eff_t-P_eff_t}
\ee
As for the threshold pressure to open the full two-channel sub-network rooted at the node at depth $t<t_1$, we check that the recursion built from Eqs.~(\ref{eq:P0and1}),(\ref{eq:P0andnot1}) is solved by
\be
P_{\text{thr}\,t}=\varepsilon^{({\cal P}_0)}_{t\,T}+\frac{T-t}{T-t_1}\left(\varepsilon^{({\cal P}_{1\hat t_1})}_{t_1\,T}-\varepsilon^{({\cal P}_0)}_{t_1\,T}\right).
\label{eq:Pthr_t_1}
\ee

The path that opens first when the pressure at the network inlet is increased from $P_0$ must obviously be a path that minimizes $P_{\text{thr}\, t=0}$. Hence the expression of the second threshold pressure $P_1$ reads
\be
P_{1}\equiv P_0+ \min_{t_1',{\cal P}'_{{1\hat t_1'}\,[t_1'+1,T]}} \frac{T}{T-t_1'}\left(\varepsilon^{({\cal P}'_{1\hat t_1'})}_{t_1'\,T}-\varepsilon^{({\cal P}_0)}_{t_1'\,T}\right).
\label{eq:Pthr_t_2}
\ee
From now on, what we shall call ${\cal P}_1$ will be a path that realizes this minimum of the threshold pressure, and $t_1$ will stand for its overlap with the path ${\cal P}_0$. Note that in formula~(\ref{eq:Pthr_t_2}), we may replace each sub-path sum $\varepsilon^{({\cal P}_{0})}_{t_1\,T},\varepsilon^{({\cal P}_{1})}_{t_1\,T}$ by the full path sum $\varepsilon^{({\cal P}_{0})}_{0\,T},\varepsilon^{({\cal P}_{1})}_{0\,T}$: the difference would remain unchanged, since ${\cal P}_0$ and ${\cal P}_1$ overlap on $[0,t_1]$.

Applying to the network a pressure $P$ chosen above $P_1$ but below the threshold pressure $P_2$ for the opening of the next channel (assuming $P_2>P_1$), the $P$-dependence of the flow reads
\be
Q_T(P)=\frac{2}{T+t_1}\left[P-\varepsilon_{0\,t_1}^{({\cal P}_0)}-\frac12\left(\varepsilon^{({\cal P}_0)}_{t_1\,T}+\varepsilon^{({\cal P}_1)}_{t_1\,T}\right)\right].
\ee
We check that this formula boils down to Eq.~(\ref{eq:Q_2(P)}) for $T=2$ and when the overlap $t_1$ is set to its only possible value in this case, namely~$1$. 

Alternatively, $Q_T(P)$ may be expressed in terms of the threshold pressures and of the overlap $t_1$:
\be
Q_T(P)=\frac{2}{T+t_1}\left(P-\frac{T+t_1}{2T}P_0-\frac{T-t_1}{2T}P_1\right).
\ee
This formula gives also the correct flow in the case so far disregarded in which the first two channels ${\cal P}_0$, ${\cal P}_1$ are degenerate and open simultaneously, namely if $P_1=P_0$.

We may use the recursions we have established in this section and generalize the calculation we have just performed for the case of one and two branches to an arbitrary number of them. Based on the same method, it is now easy to design an algorithm that automates the search for open channels in a given tree-like network, and thus the calculation of the flow. This is what we will expose now.


\subsection{Algorithm to determine the open sub-tree and the corresponding flow}
\label{sec:naive-algo}

We start with a fully-generated tree-like network of height $T$. Given this frozen disorder, we want to determine the sub-trees of open channels at any given pressure at the inlet, the thresholds $P_0,P_1,\cdots$ at which the channels open successively, as well as the flow function $Q(P)$. In this section, we shall follow closely in spirit the method exposed in Ref.~\cite{Schimmenti:2023}. 

To begin, we search for a path ${\cal P}_0$ that minimizes $\varepsilon^{({\cal P}_0)}_{0\,T}$. Let us first assume, for simplicity, that the values of $\varepsilon^{({\cal P})}_{0\,T}$ on all possible paths $\cal P$ in the tree are all distinct. Then, within this assumption, there is a single channel with lowest threshold pressure $P_0$. Once this first channel is determined, the flow parameters entering the Darcy law~(\ref{eq:Q(P)}) for this single-channel network are calculated iterating Eq.~(\ref{eq:flow-parameters-1}) from the leaves to the root.

To find the next channels to open and the next threshold pressure $P_1$, we loop over all nodes of the path ${\cal P}_0$, namely over the depth $t_1$, with $0<t_1<T$. For each $t_1$, we search for the path ${\cal P}_{1}$ that minimizes $\varepsilon_{t_1+1\,T}^{({\cal P}_{1})}$. Then, using Eqs.~(\ref{eq:P0and1}) and~(\ref{eq:P0andnot1}), we compute the minimum pressure $P_{\text{thr}\,0}$ that has to be applied at the inlet of the full network to open the second channel at the outlet of this very node, assuming that all other closed channels of the tree remain closed. We eventually open the channel that minimizes this threshold pressure. We arrive at a two-branch subtree, such as the one singled out in Fig.~\ref{fig:Cayley-tree}. The flow parameters are subsequently computed by iterating Eqs.~(\ref{eq:flow-parameters-1}) and~(\ref{eq:flow-parameters-2}) going up the tree, from the leaves to the root.

We repeat the steps just described to the two-branch sub-tree we have determined in order to search for the next channel to open. We test the opening pressure of all channels attaching to that sub-tree. These threshold pressures are determined iterating Eqs.~(\ref{eq:P0and1}) and~(\ref{eq:P0andnot1}). We open the channel endowed with the lowest threshold and compute its flow parameters.

We iterate this overall procedure for the next branches, until some predefined stopping condition is satisfied, depending on the observable we aim at measuring. We see that in this way, we are eventually able to compute the flow $Q_T(P)$ for any value of $P$.

In the case in which there are $n$ channels that have the same threshold pressure, the algorithm can be kept unchanged: we just open these degenerate channels successively, in random order to each other.


\paragraph{Reduction to the directed polymer problem.}

This algorithm can be trivially modified to solve the problem of the determination of the successive configurations of the directed polymer problem, from the ground state to higher excited states. Instead of computing the threshold pressures $P_0$, $P_1$, $P_2,\cdots$ and opening the channels in increasing order of these thresholds, we order the directed paths ${\cal P}_0$, ${\cal P}_1$, ${\cal P}_2\cdots$ on the tree in increasing sequence $\varepsilon_0\equiv\varepsilon^{({\cal P}_0)}_{0\,T}$, $\varepsilon_1\equiv\varepsilon^{({\cal P}_1)}_{0\,T}$, $\varepsilon_2\equiv\varepsilon^{({\cal P}_2)}_{0\,T}\cdots$ of the path sums of the $\tau$'s. The latter quantities correspond to the energies of the directed polymer, and the associated paths ${\cal P}_0$, ${\cal P}_1$, ${\cal P}_2,\cdots$ represent the corresponding polymer configurations.


\section{Optimized algorithm for successive channel opening}
\label{sec:advanced-algo}

The algorithm just described requires the prior generation of the network. The complexity of the generation of a tree of height $T$ grows like the number of bonds, namely exponentially in $T$. This definitely restricts the values of $T$ that may be reached, in any practical implementation, to a few dozen units. With such an algorithm, whatever the computing power available, it will never be possible to increase $T$ by more than a factor of order~1.

However, there is a much more efficient way to proceed. Instead of generating all elementary thresholds $\tau$ a priori, we may grow the trees branch-by-branch. The new algorithm we shall introduce is inspired from that described in Ref.~\cite{Brunet:2020yiq}, which was designed to generate exactly the tip region of a branching random walk at large times. The initial motivation came from particle physics: the problem there was to be able to efficiently generate scattering configurations relevant to electron-nucleus collisions, in some suitable asymptotic high-energy regime. (The interested reader is referred to~\cite{Angelopoulou:2023qdm} for a recent review on the connections between scattering in particle physics and general branching processes). But the algorithm may also be used to generate the lowest-lying energy levels of a directed polymer in a random medium in the mean-field approximation, which turns out to be a problem in the same class. Notably, the underlying method can be interpreted as a spinal decomposition of branching random walks, a technique previously recognized in mathematical literature (see e.g. Refs.~\cite{Hardy:2009,Harris:2017}) but not previously developed into an algorithm.

Let us start by describing the algorithm to solve the directed polymer problem, before explaining how to adapt it to address the Darcy problem.


\subsection{Energy-ordered tree generation}

Here, we explain how to sequentially construct the configurations $\cal P$ of increasing energy $\varepsilon^{({\cal P})}_{0\,T}$, where $\varepsilon^{({\cal P})}_{t_1\,t_2}$ was defined in Eq.~(\ref{eq:def-varepsilon-path}), starting from the ground state.

The main idea is that the distribution of the ground state energy, $\min_{\cal P}\varepsilon_{0\,T}^{({\cal P})}\equiv \varepsilon_0$, is determined by a non-linear evolution equation that is straightforward to solve numerically (at least when all energies are discrete). This allows us to determine the ground state energy without generating the entire random tree. Once $\varepsilon_0$ is known, a polymer configuration ${\cal P}_0$ with that minimum energy can be exactly constructed using a branching random walk defined by two elementary processes with non-trivial but definite probabilities. Furthermore, we consider the first excited state as the configuration that minimizes the ground state energy among all polymer configurations branching off ${\cal P}_0$. Its energy can also be determined without needing to generate all configurations a priori. This process can be iterated until all relevant configurations have been generated.

We will successively discuss the probability distribution of the ground state energy, the construction of the ground state configuration(s), and then the excited states.


\paragraph{Ground state energy.}

We denote by $\P_T(\varepsilon_0>\varepsilon)$ the probability that the energy of the ground state for a tree of height $T$ is greater than $\varepsilon$. We call $p(\tau)$ the distribution of the random energies $\tau$ on the bonds. Then
\be
\P_T(\varepsilon_0>\varepsilon)=\sum_{\tau_{01}}p(\tau_{01})\P_{T-1}(\varepsilon^\text{bin}_0>\varepsilon-\tau_{01})\,,
\label{eq:PT}
\ee
where $\P_s(\varepsilon_0^\text{bin}>\varepsilon')$ stands for the probability that the energy of the ground state of a perfect binary tree of height~$s$ is larger than $\varepsilon'$. The discrete sum becomes an integral if $p$ is a probability density. But for practical reasons, we will assume that $\tau$ takes only integer values. Other cases may be recovered through an appropriate scaling, and possible limiting procedures.

The distribution of $\varepsilon_0^\text{bin}$ obeys a recursion relation in the height of the binary tree:
\be
\P_{s+1}(\varepsilon_0^\text{bin}>\varepsilon')=\left(\sum_\tau p(\tau)\,\P_s(\varepsilon_0^\text{bin}>\varepsilon'-\tau)\right)^2\,.
\label{eq:evol-P}
\ee
The ground state energy of a polymer made of a single monomer is of course null, and thus the initial condition simply reads
\be
\P_{s=0}(\varepsilon_0^\text{bin}>\varepsilon')={\mathbbm 1}_{\{\varepsilon'<0\}}\,.
\ee
The derivation of recursions such as Eq.~(\ref{eq:evol-P}) is standard (see e.g. \cite{Derrida:2016} for the discussion of a very similar model).

It proves convenient to introduce the probability that the ground state energy $\varepsilon_0^\text{bin}$ is not larger than some $\varepsilon'$. It is just the complementary of the previous probability:
\be
u_s(\varepsilon')\equiv \P_{s}(\varepsilon_0^\text{bin}\leq \varepsilon')=1-\P_{s}(\varepsilon_0^\text{bin}>\varepsilon').
\label{eq:def-u}
\ee
From Eq.~(\ref{eq:evol-P}), we see that it obeys the non-linear finite-difference equation
\be
u_{s+1}(\varepsilon')=2\sum_\tau p(\tau)u_s(\varepsilon'-\tau)-\left(\sum_\tau p(\tau)u_s(\varepsilon'-\tau)\right)^2,\quad
\text{with}\ u_{s=0}(\varepsilon')={\mathbbm 1}_{\{\varepsilon'\geq 0\}}.
\label{eq:FKPP}
\ee
Such equations can be shown to admit ``pulled-front solutions'', and in this sense, belong to the universality class of the Fisher-Kolmogorov-Petrovsky-Piscounov (FKPP) equation \cite{Fisher:1937,KPP:1937} (see Ref.~\cite{VanSaarloos:2003} for an extensive review, as well as Appendix~\ref{sec:TW}).

From Eqs.~(\ref{eq:def-u}) and~(\ref{eq:PT}), the distribution of $\varepsilon_0$ is found to read
\be
\P_T(\varepsilon_0=\varepsilon)=\sum_{\tau_{01}}p(\tau_{01})
\left[u_{T-1}(\varepsilon-\tau_{01})-u_{T-1}(\varepsilon-\tau_{01}-1)\right].
\ee


\paragraph{Building one configuration of minimal energy.}

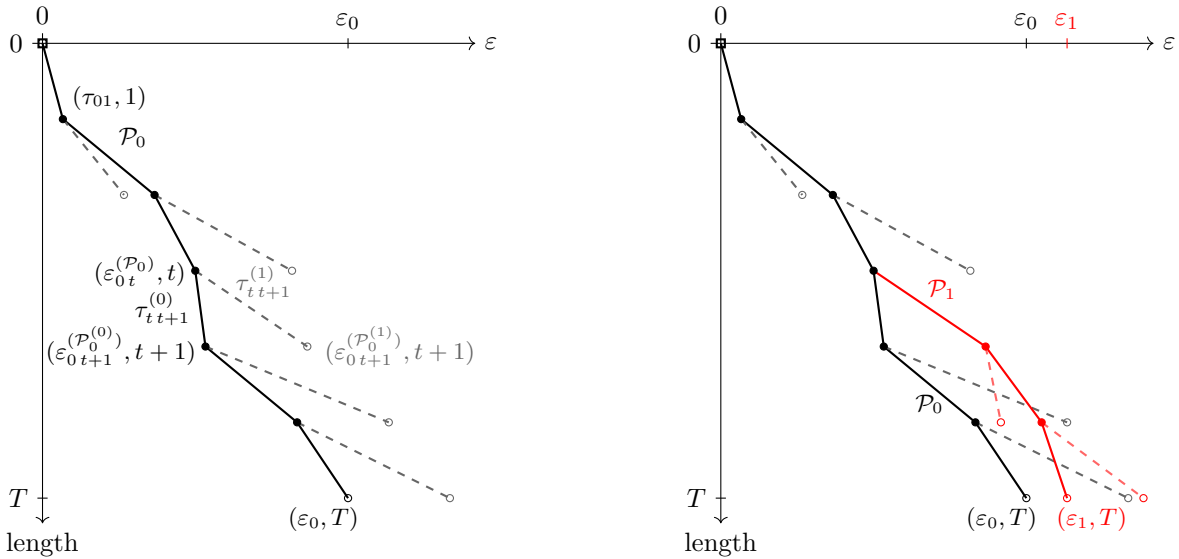
\begin{figure}[t]
  \begin{subfigure}[t]{0.45\textwidth}
    \centering\begin{tikzpicture}[scale=0.67]
        \draw[->] (0.,0) -- (0.,-9.5) node[anchor=north] {length};
        \draw (-0.2,0) node[anchor=east] {$0$} -- (0,0);
        \draw (-0.1,-9) node[anchor=east] {$T$} -- (0.1,-9);
        \draw[->] (0,0) -- (8.5,0) node[anchor=west] {$\varepsilon$};
        \draw (0,0.2) node[anchor=south] {$0$} -- (0,0);
        \draw (6,0.1) node[anchor=south] {$\varepsilon_0$} -- (6,-0.1);
        \coordinate (V0) at (0,0);
        \coordinate (V1) at (0.4,-1.5);
        \coordinate (V2) at (2.2,-3);
        \coordinate (W2) at (1.6,-3);
        \coordinate (V3) at (3.0,-4.5);
        \coordinate (W3) at (4.9,-4.5);
        \coordinate (V4) at (3.2,-6);
        \coordinate (W4) at (5.2,-6);
        \coordinate (V5) at (5,-7.5);
        \coordinate (W5) at (6.8,-7.5);
        \coordinate (V6) at (6,-9);
        \coordinate (W6) at (8,-9);
        \draw[black,thick] (V0) node[draw,rectangle,inner sep=0pt,minimum size=3pt,fill=none]{} -- (V1) -- node[anchor=south west] {\small ${\cal P}_0$} (V2) -- (V3) -- node[anchor=east] {\small $\tau^{(0)}_{t\,t+1}$} (V4) -- (V5) -- (V6);
        \draw[black!60,thick,dashed] (V1) -- (W2);
        \draw[black!60,thick,dashed] (V2) -- (W3);
        \draw[black!60,thick,dashed] (V3) -- node[anchor=south] {\small $\quad \tau^{(1)}_{t\,t+1}$}  (W4);
        \draw[black!60,thick,dashed] (V4) -- (W5);
        \draw[black!60,thick,dashed] (V5) -- (W6);
        
        \node[anchor=south west] at (V1) {\small $(\tau_{01},1)$};
        \node[anchor=east] at (V3) {\small $(\varepsilon^{({\cal P}_0)}_{0\,t},t)$};
        \node[anchor=east] at (V4) {\small $(\varepsilon^{({\cal P}_0^{(0)})}_{0\,t+1},t+1)$};
        \node[black!60,anchor=west] at (W4) {\small $\ (\varepsilon^{({\cal P}_0^{(1)})}_{0\,t+1},t+1)$};
        \node[anchor=north] at (V6) {\small $(\varepsilon_0,T)\quad\quad$};

        \filldraw[black] (V1) circle (2pt);
        \filldraw[black] (V2) circle (2pt);
        \draw[black!60] (W2) circle (2pt);
        \filldraw[black] (V3) circle (2pt);
        \draw[black!60] (W3) circle (2pt);
        \filldraw[black] (V4) circle (2pt);
        \draw[black!60] (W4) circle (2pt);
        \filldraw[black] (V5) circle (2pt);
        \draw[black!60] (W5) circle (2pt);
        \draw[black] (V6) circle (2pt);
        \draw[black!60] (W6) circle (2pt);
\end{tikzpicture}
    \caption{Construction of a ground state configuration ${\cal P}_0$ of the directed polymer.}
    \label{fig:first-open-channel}
  \end{subfigure}
  \hfill  
  \begin{subfigure}[t]{0.5\textwidth}
    \centering\begin{tikzpicture}[scale=0.67]
        \draw[->] (0.,0) -- (0.,-9.5) node[anchor=north] {length};
        \draw (-0.2,0) node[anchor=east] {$0$} -- (0,0);
        \draw (-0.1,-9) node[anchor=east] {$T$} -- (0.1,-9);
        \draw[->] (0,0) -- (8.5,0) node[anchor=west] {$\varepsilon$};
        \draw (0,0.2) node[anchor=south] {$0$} -- (0,0);
        \draw (6,0.1) node[anchor=south] {$\varepsilon_0$} -- (6,-0.1);
        \draw[red] (6.8,0.1) node[anchor=south] {$\varepsilon_1$} -- (6.8,-0.1);
        \coordinate (V0) at (0,0);
        \coordinate (V1) at (0.4,-1.5);
        \coordinate (V2) at (2.2,-3);
        \coordinate (W2) at (1.6,-3);
        \coordinate (V3) at (3.0,-4.5);
        \coordinate (W3) at (4.9,-4.5);
        \coordinate (V4) at (3.2,-6);
        \coordinate (W4) at (5.2,-6);
        \coordinate (V5) at (5,-7.5);
        \coordinate (W5) at (6.8,-7.5);
        \coordinate (X5) at (6.3,-7.5);
        \coordinate (Y5) at (5.5,-7.5);
        \coordinate (V6) at (6,-9);
        \coordinate (W6) at (8.,-9);
        \coordinate (X6) at (6.8,-9);
        \coordinate (Y6) at (8.3,-9);
        \draw[black,thick] (V0) node[draw,rectangle,inner sep=0pt,minimum size=3pt,fill=none]{} -- (V1) -- (V2) -- (V3) -- (V4) -- node[anchor=north] {\small ${\cal P}_0$} (V5) -- (V6);
        \draw[black!60,thick,dashed] (V1) -- (W2);
        \draw[black!60,thick,dashed] (V2) -- (W3);
        \draw[red,thick] (V3) -- node[anchor=south] {\small $\quad {\cal P}_1$}  (W4);
        \draw[black!60,thick,dashed] (V4) -- (W5);
        \draw[red,thick] (W4) -- (X5);
        \draw[red!60,thick,dashed] (W4) -- (Y5);
        \draw[black!60,thick,dashed] (V5) -- (W6);
        \draw[red,thick] (X5) -- (X6);
        \draw[red!60,thick,dashed] (X5) -- (Y6);
 
        \node[anchor=north] at (V6) {\small $(\varepsilon_0,T)\quad\quad$};
        \node[red,anchor=north] at (X6) {\small $\quad\quad (\varepsilon_1,T)$};

        \filldraw[black] (V1) circle (2pt);
        \filldraw[black] (V2) circle (2pt);
        \draw[black!60] (W2) circle (2pt);
        \filldraw[black] (V3) circle (2pt);
        \draw[black!60] (W3) circle (2pt);
        \filldraw[black] (V4) circle (2pt);
        \filldraw[red] (W4) circle (2pt);
        \filldraw[black] (V5) circle (2pt);
        \draw[black!60] (W5) circle (2pt);
        \filldraw[red] (X5) circle (2pt);
        \draw[red] (Y5) circle (2pt);
        \draw[black] (V6) circle (2pt);
        \draw[black!60] (W6) circle (2pt);
        \draw[red] (X6) circle (2pt);
        \draw[red] (Y6) circle (2pt);

\end{tikzpicture}
    \caption{The ground state configuration ${\cal P}_0$ and the first excited state ${\cal P}_1$, in the case in which the energy levels are not degenerate.}
    \label{fig:second-open-channel}
  \end{subfigure}
  \caption{Illustration of the construction of the lowest-lying states of a polymer. The first monomer is located at coordinates $(0,0)$ in this plane, and is marked by a square. Each black dot marks a node occupied by a monomer in the binary configuration tree. The ordinate of the latter, denoted by $t$, corresponds to its rank in the sequence making up the polymer. Its abscissa $\varepsilon$ is the energy of the configuration of the sub-polymer made of the first $t$ monomers. {\it Full lines}: bonds that belong to a configuration below a given energy. {\it Dashed lines}: bonds that belong to higher-energy configurations. This sketch is easily taken over to the Darcy problem: the polymer length becomes the network depth, and one substitutes $\varepsilon_{0},\varepsilon_1$ with $P_{0},P_1$.}
\end{figure}

Once the energy $\tau_{01}$ of the first bond and the minimum energy $\varepsilon_0$ are known, the paths ${\cal P}_0\equiv{\cal P}_{0\,[0,T]}$ in the tree corresponding to the minimal-energy configurations can be constructed iteratively, using a biased branching random walk that we shall now fully specify.

We shall assume that the sub-path ${\cal P}_{0\,[0,t]}$ of a configuration of minimum energy (which is not necessarily unique) is known. Given this, we can determine the law governing the next bond energy $\tau_{t\,t+1}$. The sub-path ${\cal P}_{0\,[t,T]}$, which we seek to construct, corresponds to a minimal-energy configuration of a polymer with $T-t+1$ monomers. This sub-path exists within the configuration space represented by the binary tree rooted in the node of ${\cal P}_{0\,[0,t]}$ at depth $t$. The energy of this sub-path is fixed to $\varepsilon^{({\cal P}_0)}_{t\,T}= \varepsilon_0-\varepsilon^{({\cal P}_0)}_{0\,t}$.

We now proceed with the construction. The sub-polymer ${\cal P}_{0\,[0,t]}$ has two possible continuation bonds, with energies denoted by $\tau^{(0)}_{t\,t+1}$ and $\tau^{(1)}_{t\,t+1}$, respectively (see Fig.~\ref{fig:first-open-channel}). To determine their joint distribution, we need the probabilities $R_t^{\varepsilon_0}(\varepsilon)$ and $B_t^{\varepsilon_0}(\varepsilon)$ which describe the likelihood that a polymer configuration with a segment energy $\varepsilon$ over $[0,t]$ will have a minimum total energy either equal to $\varepsilon_0$ or greater than $\varepsilon_0$, respectively. These probabilities correspond to the likelihood that the ground state energy of sub-polymers rooted at $(\varepsilon,t)$, and made of $T-t+1$ monomers, is equal to or greater than $\varepsilon_0-\varepsilon$. It is then enough to recall that $u_s(\varepsilon')$ is the probability that the ground state energy of a binary tree of height $s$ is not larger than $\varepsilon'$. The sought probabilities follow immediately:
\be
R_t^{\varepsilon_0}(\varepsilon)=u_{T-t}(\varepsilon_0-\varepsilon)-u_{T-t}(\varepsilon_0-\varepsilon-1)
\quad\text{and}\quad
B_t^{\varepsilon_0}(\varepsilon)=1-u_{T-t}(\varepsilon_0-\varepsilon)\,.
\label{eq:RandB}
\ee

Let us now introduce the paths ${\cal P}_0^{(0)}$ and ${\cal P}_0^{(1)}$, defined such that they both overlap with ${\cal P}_0$ on the segment $[0,t]$ and then follow the two possible outgoing bonds from the node of ${\cal P}_0$ located at depth $t$:
\be
{\cal P}_{0\,[0,t]}^{(0)}={\cal P}_{0\,[0,t]}^{(1)}={\cal P}_{0\,[0,t]},
\quad\text{and}\quad 
\varepsilon^{({\cal P}_0^{(0)})}_{t\,t+1}=\tau_{t\,t+1}^{(0)},
\ \varepsilon^{({\cal P}_0^{(1)})}_{t\,t+1}=\tau_{t\,t+1}^{(1)}.
\ee
The labels are chosen so that ${\cal P}_0^{(0)}$ will eventually coincide with ${\cal P}_0$. We must also impose the following conditions on the paths ${\cal P}_0^{(0)}$ and ${\cal P}_{0}^{(1)}$:
\be
\min_{{\cal P}_{0\,[t+1,T]}^{(0)}}\varepsilon^{({\cal P}_0^{(0)})}_{0\,T}=\varepsilon_0,
\quad\text{and}\quad
\min_{{\cal P}_{0\,[t+1,T]}^{(1)}}\varepsilon^{({\cal P}_0^{(1)})}_{0\,T}\geq \varepsilon_0.
\label{eq:constraints-follow-up-paths}
\ee
We are now ready to express the probability distribution for the pair $(\tau^{(0)}_{t\,t+1},\tau^{(1)}_{t\,t+1})$.

There are two cases to distinguish, depending on whether the inequality in Eq.~(\ref{eq:constraints-follow-up-paths}) holds as an equality, or as a strict inequality:
\begin{enumerate}[label=(\roman*)]
\item\label{case:degenerate}
{\em Equality:} the possible configurations ${\cal P}_0^{(0)}$ and ${\cal P}_0^{(1)}$ have the same minimum energy $\varepsilon_0$.  Then, the joint probability of the latter event and that the bond energies have the definite values $\tau$, $\tau'$, given that the minimal energy of the overall polymer configuration ${\cal P}_0$ is $\varepsilon_0$, reads
  \begin{multline}  
  \P\bigg(\tau^{(0)}_{t\,t+1}=\tau,
  \tau^{(1)}_{t\,t+1}=\tau';\min_{{\cal P}_{0\,[t+1,T]}^{(0)}}\varepsilon^{({\cal P}_0^{(0)})}_{0\,T}=\min_{{\cal P}_{0\,[t+1,T]}^{(1)}}\varepsilon^{({\cal P}_0^{(1)})}_{0\,T}
  =\varepsilon_0\bigg|\min_{{\cal P}_{0[t,T]}}\varepsilon^{({\cal P}_{0})}_{0\,T}=\varepsilon_0\bigg)\\
  =\frac{p(\tau)p(\tau')R_{t+1}^{\varepsilon_0}(\varepsilon^{({\cal P}_0)}_{0\,t}+\tau)
    R_{t+1}^{\varepsilon_0}(\varepsilon^{({\cal P}_0)}_{0\,t}+\tau')}
       {R_t^{\varepsilon_0}(\varepsilon^{({\cal P}_0)}_{0\,t})}.
       \label{eq:proba-degenerate}
  \end{multline}
   
\item\label{case:non-degenerate}
  {\em Strict inequality:} one of the possible configurations, which we chose to label as ${\cal P}_0^{(1)}$, has ground state energy larger than $\varepsilon_0$. In this case,
  \begin{multline}
    \P\bigg(\tau^{(0)}_{t\,t+1}=\tau,
    \tau^{(1)}_{t\,t+1}=\tau';\min_{{\cal P}_{0\,[t+1,T]}^{(0)}}\varepsilon^{({\cal P}_0^{(0)})}_{0\,T}=\varepsilon_0,
    \min_{{\cal P}_{0\,[t+1,T]}^{(1)}}\varepsilon^{({\cal P}_0^{(1)})}_{0\,T}
    >\varepsilon_0\bigg|\min_{{\cal P}_{0[t,T]}}\varepsilon^{({\cal P}_{0})}_{0\,T}=\varepsilon_0\bigg)\\
    =2\frac{p(\tau)p(\tau')R_{t+1}^{\varepsilon_0}(\varepsilon^{({\cal P}_0)}_{0\,t}+\tau)
      B_{t+1}^{\varepsilon_0}(\varepsilon^{({\cal P}_0)}_{0\,t}+\tau')}
    {R_t^{\varepsilon_0}(\varepsilon^{({\cal P}_0)}_{0\,t})}\,.
    \label{eq:proba-nondegenerate}
  \end{multline}
  
\end{enumerate}
We check that the sum of these probabilities marginalized with respect to $\tau$ and $\tau'$ is unitary, as it should: it is enough to replace $R$ and $B$ appearing in the above expressions by their definitions~(\ref{eq:RandB}) in terms of $u$'s and perform the sums, using the evolution equation~(\ref{eq:FKPP}).

We first determine if we are in case~\ref{case:degenerate} or~\ref{case:non-degenerate} using the probabilities~(\ref{eq:proba-degenerate}),(\ref{eq:proba-nondegenerate}) summed over $\tau$ and $\tau'$. In a second step, we draw $\tau$ and $\tau'$ from either the probability in Eq.~(\ref{eq:proba-degenerate}) or that in Eq.~(\ref{eq:proba-nondegenerate}), with appropriate normalization following from the conditioning to be in cases~\ref{case:degenerate} or \ref{case:non-degenerate} respectively.

In both cases, the outgoing bond labeled $(0)$ extends the path ${\cal P}_0$. If one ends up in the non-degenerate case~\ref{case:non-degenerate}, then once $\tau_{t\,t+1}^{(1)}$ is determined, one draws the minimal energy that can have the other polymer configuration: the next configuration to be constructed will be one among those realizing this minimum. The distribution of $\min_{{\cal P}_0^{(1)}}\varepsilon^{({\cal P}_0^{(1)})}_{t+1\,T}$ reads
\be
\P\left(\bigg.\min_{{\cal P}_{0\,[t+1,T]}^{(1)}}\varepsilon^{({\cal P}_0^{(1)})}_{t+1\,T}=\varepsilon\bigg|\min_{{\cal P}_{0\,[t+1,T]}^{(1)}}\varepsilon^{({\cal P}_{0}^{(1)})}_{0\,T}>\varepsilon_0\right)
=\frac{u_{T-t-1}(\varepsilon)-u_{T-t-1}(\varepsilon-1)}{B^{\varepsilon_0}_{t+1}(\varepsilon^{({\cal P}_0)}_{0\,t}+\tau_{t\,t+1}^{(1)})}
{\mathbbm 1}_{\{\varepsilon>\varepsilon_0-\varepsilon^{({\cal P}_0)}_{0\,t}-\tau_{t\,t+1}^{(1)}\}}\,.
\ee
We shall call ${\cal P}_{1\hat t}$ a polymer configuration that realizes this minimum. Note that at this point, we don't need to have fully constructed this configuration: all we need for the next step is the value of its energy, and we have just shown that it can be drawn independently of knowledge of the detailed configuration.

One iterates this construction of the segments of increasing depth until the first path ${\cal P}_0$ is complete. Whether there are degeneracies or not, we shall always build the configuration labeled $(0)$ first, leaving the construction of the configuration $(1)$ for a next step that we shall now describe. We store $\tau^{(1)}_{t\,t+1}$, and if there is no degeneracy, namely in case \ref{case:non-degenerate}, we also store the values of the minimum energy of the polymer configurations branching off this node of ${\cal P}_0$.


\paragraph{Next configurations.}

Once a ground state configuration has been constructed, we search for the next lowest energy level. There may be several configurations having the same ground state energy. In this case, we just pick randomly one of the nodes off which two sub-polymer configurations with degenerate energies branch, say at depth $t$, and we construct the second polymer configuration. The latter results from a branching random walk defined by the probabilities~(\ref{eq:proba-degenerate}) and~(\ref{eq:proba-nondegenerate}). We eventually get a path ${\cal P}_0'$, with overlap $t$ with ${\cal P}_0$. In the same way as we did for the path ${\cal P}_0$, we draw and store the bond energies of the polymer configurations which branch off the newly-constructed nodes of ${\cal P}_0'$, as well as the minimum energies of these configurations.

If there is no degeneracy of the ground-state energy $\varepsilon_0$, or after having opened all degenerate paths, we determine the energy of the first excited state. It must be the lowest energy of all configurations ${\cal P}_{1\hat t'}$, where $t'$ runs over all nodes of the path ${\cal P}_0$ (or of the set of paths ${\cal P}_0,{\cal P}'_0,\cdots$ of energy $\varepsilon_0$). In the non-degenerate case, in which the sub-tree of configurations of energy $\varepsilon_0$ consists of a single branch,
\be
\varepsilon_1=\min_{t'}\left(\varepsilon_{0\,t'}^{({\cal P}_0)}+\varepsilon_{t'\,T}^{({\cal P}_{1\hat t'})}\right).
\label{eq:next-to-minimum-energy}
\ee
The polymer configuration ${\cal P}_1$ that branches off ${\cal P}_0$ at the node labeled, say, $t$, and minimizes this energy is then constructed, again from a branching random walk. The latter is defined by the probabilities~(\ref{eq:proba-degenerate}) and~(\ref{eq:proba-nondegenerate}), up to the replacements ${\cal P}_0\to{\cal P}_1$ (as well as for the super-scripted paths with $(0)$ and $(1)$) and $\varepsilon_0\to \varepsilon_1$. A sketch of two configurations corresponding to the two lowest values of the energy in the non-degenerate case is displayed in Fig.~\ref{fig:second-open-channel}.

In the degenerate case, the minimum in Eq.~(\ref{eq:next-to-minimum-energy}) is taken over all nodes of the sub-tree of the configurations of energy $\varepsilon_0$, and in the term $\varepsilon_{0\,t'}^{({\cal P}_0)}$, the path ${\cal P}_0$ is replaced by a path of that tree that includes the considered node.

The whole procedure is thus iterated until some stopping condition is reached, e.g. on the maximum energy or on the number of configurations.


\subsection{Threshold pressure-ordered tree generation}

The algorithm to open the successive channels as the pressure at the inlet is increased can almost be taken over from the one just described for the directed polymer problem. The only variation will turn out to be in the ordering of the channels to open.

Let us start by recalling the dictionary between the directed polymer problem of finding the lowest-lying energy configurations, and the Darcy problem of opening channels when the pressure at the inlet of the network is dialed up. The bonds between monomers correspond to throats which join at the trivalent nodes of the tree-like network. The latter are identified to pores in the Darcy problem. The rank $t$ of the monomers in the polymer is the depth of the throat joints, its size $T$ is the height of the network. The bond energies $\tau$ are the elementary threshold pressure difference of the considered throats. 

However, in the DP problem, the successive configurations that we generate are ordered in energy $\varepsilon_0,\varepsilon_1,\cdots$. In the Darcy problem instead, we open the channels in increasing order of the applied pressure thresholds $P_0, P_1,\cdots$. In general, the latter do not correspond to the successive energies $\varepsilon_{0\,T}^{({\cal P}_0)}, \varepsilon_{0\,T}^{({\cal P}_1)},\cdots$.

To construct the sub-tree of open channels, we start with the first channel, which matches exactly a configuration that possesses the ground state energy in the DP case. We then search for the channels that open successively as the pressure is increased. The procedure to find the next channel, after ${\cal P}_0$, to open is exactly the same as the one implemented in the naive algorithm in Sec.~\ref{sec:naive-algo}. Given the open sub-tree, instead of looking for the next minimum energy, we compute the threshold pressure at the inlet for each of the closed channels attaching to it, assuming that all other channels remain closed. Instead of the energy $\varepsilon_1$ given by Eq.~(\ref{eq:next-to-minimum-energy}), the relevant variable $P_1$ is this minimum threshold pressure~(\ref{eq:Pthr_t_2}). We construct and open one of the channels that possess the threshold pressure $P_1$. 

We then iterate, using the more general equations~(\ref{eq:P0and1}),(\ref{eq:P0andnot1}) in order to compute $P_2$,$P_3\cdots$, until some stopping condition is satisfied.


\paragraph{Performance of the new algorithm.}

The major improvement brought by our new method is obviously that we do not need to generate the full tree of throats (i.e. the random medium) a priori. We are able to generate the branches that correspond exactly to the channels that open, in successive order when the exerted pressure is increased. This is all the more useful as we only need a number of open channels that is much smaller than the total number $2^{T-1}$ of possible channels.

Consequently, the complexity of our new algorithm is linear in the height $T$ of the tree-like network, instead of exponential as for the more straightforward algorithm described in Sec.~\ref{sec:naive-algo}. At each threshold, the calculation time is dominated by the time needed to search for the next channels to open. Hence for a total number of channels $n^\text{ch}$, the complexity is bounded by ${\cal O}\left(T\times \left({n^{\text{ch}}}\right)^2\right)$. This allows us to increase the height accessible to numerical investigations by several orders of magnitude.

Note however that a solution $u_t(\varepsilon')$ of Eq.~(\ref{eq:FKPP}) needs to be pre-computed and stored for all $t$ in $[0,T]$, and for all relevant values of $\varepsilon'$, the number of which becomes proportional to $\sqrt{T}$ at large~$T$. But in practice, this does not impose any major restriction on the calculations that are useful for the insight we wish to gain from a numerical study.

All numerical data presented and discussed below were generated running our new algorithm on an off-the-shelf laptop computer, for not more than a few hours for each parameter setting. Note that we also implemented the naive algorithm, and performed accurate comparisons of the calculations of observables using both algorithms for low values of $T$, in order to validate our implementation.


\section{Analysis of the model}
\label{sec:results}

In this section, we apply our optimized algorithm to generate new relevant numerical data. Before presenting our results, we will briefly revisit a few conjectures proposed in Ref.~\cite{Schimmenti:2023} (with detailed derivations in Ref.~\cite{Schimmenti:2023_supplemental}). These conjectures lead to asymptotic formulas that our data will allow us to test, thereby providing an opportunity to confirm or refute the underlying theoretical framework.

\subsection{Analytical formulas in two asymptotic limits}

According to Ref.~\cite{Schimmenti:2023}, there are essentially two limits in which the flow can be worked out: (i) when the pressure $P$ exerted at the inlet is large enough so that all channels are open, (ii) for fixed~$P$, but very large tree height~$T$. Let us consider these limits in turn.


\paragraph{High inlet pressure.}

We first assume that $P$ is set above the threshold for the opening of the last of the $2^{T-1}$ channels. Then the tree of height $T$ is complete, as well as all its sub-trees. Looking at equation~(\ref{eq:flow-parameters-2}) for $\kappa_\text{eff}$, it is easy to convince oneself that the latter depends only on the depth of the node one considers. It depends neither on the $\tau$'s, nor on the particular node at a given depth.

Let us call $\kappa_t$ the value of $\kappa_\text{eff}$ at the top of a (sub-)tree of generic height $t$, and $P_t^*$ the value of the effective threshold pressure at the inlet of the same (sub-)tree. [Note that at variance with $\kappa_t$, $P_t^*$ is a random variable also in this case in which all channels are open, meaning that it depends on the particular realization of the (sub-)tree]. We then apply Eq.~(\ref{eq:flow-parameters-2}) to the node of lowest depth, namely at the top of the tree: \be
\kappa_T=\frac{2\kappa_{T-1}}{1+2\kappa_{T-1}}
\quad\text{and}\quad
P_{T}^{*}=\tau_{01}+\frac12\left(P_{T-1}^{*(0)}+P_{T-1}^{*(1)}\right)\,.
\ee
The equation for $\kappa_T$ is a recursion. With the initial condition $\kappa_1=1$, it can be easily solved:
\be
\kappa_T=\frac{2^{T-1}}{2^T-1}\,.
\ee
The equation for $P_T^*$ can also be turned into a recursion by taking the average over realizations of the disorder. The obtained equation for $\overline{P^*_T}$ can be trivial solved, given the initial condition $\overline{P^*_0}=P^*_0=0$:
\be
\overline{P^*_T}=T\times\overline{\tau}\,.
\ee
Hence these simple calculations yield the following useful asymptotical expressions:
\be
\kappa_\text{eff}\to\frac12
\quad\text{and}\quad
\frac{\overline{P^*_\text{eff}}}{T}\to\overline{\tau}\quad\text{when $P\to\infty$}.
\label{eq:asymptotics-kappa_eff-P_eff}
\ee
Note that $P^*_\text{eff}$ is a self-averaging quantity, so we expect that at large pressure $P$ and large tree height $T$, all the flow curves converge to 
\begin{equation}
Q_T(P)=\frac12\left(P-\overline{\tau} T\right).
\label{eq:asymptotics-flow}
\end{equation}

\paragraph{Great tree height.}

When the pressure at the inlet is finite, a finite number of channels are open. Of course, the channels that are most likely to open when the pressure at the inlet increases are expected to sit at low $\varepsilon_{0\,T}$. But it is also advantageous that they have a small overlap with each other, namely that the paths get quickly separated. This is easily seen from the minimization of $P_{\text{thr}\,0}$ in Eq.~(\ref{eq:Pthr_t_1}), realized in Eq.~(\ref{eq:Pthr_t_2}) by the paths ${\cal P}_0$ and ${\cal P}_1$:
\be
P_{1}=P_0+\frac{T}{T-t_1}\left(\varepsilon_{0\,T}^{({\cal P}_{1})}-\varepsilon_{0\,T}^{({\cal P}_{0})}\right).
\label{eq:P1}
\ee
To get the threshold pressure $P_1$ as close as possible to $P_0$, it is apparent that the path sum $\varepsilon_{0\,T}^{({\cal P}_{1})}$ of the newly-opened channel needs to be close to that of the first channel $\varepsilon_{0\,T}^{({\cal P}_{0})}$, and at the same time, the overlap $t_1$ of the two channels needs to be as small as possible. Consequently, we may think that the $\varepsilon$'s of the open channels are essentially independent identically-distributed random variables, like the energy levels in the Random Energy Model (REM) \cite{Derrida:1980} (see Appendix~\ref{sec:Poisson-REM}). 

From this observation, one predicts a definite form for the distribution of the number of open channels in a given pressure range above $P_0$, see Eq.~(\ref{eq:Poisson-REM}), that can be checked with the help of our numerical code. The mean flow for a pressure $P$ such that $P-P_0=x$, where $x$ is kept fixed, can also be evaluated in this simplified picture~\cite{Schimmenti:2023}: 
\be
\overline{Q_T(P_0+x)}=\frac{e^{\beta_c x}-1}{\beta_c T}\,.
\label{eq:Q-large-T}
\ee

Let us give an alternative proof of this formula to that in Ref.~\cite{Schimmenti:2023}. For pressures different from the opening pressures of the various channels, the flow $Q$ is the affine function of the pressure as described in Eq.~(\ref{eq:Q(P)}). Hence its derivative is simply the piece-wise constant function $\kappa_\text{eff}$:
\be
\frac{dQ_T(P_0+x)}{dx}=\kappa_\text{eff}(P_0+x).
\label{eq:derivative-Q}
\ee
On the other hand, it is easy to see, iterating Eq.~(\ref{eq:kappa_eff_t-P_eff_t}), that for a deep tree made of $n^\text{ch}$ branches which have negligible overlaps:
\be
\kappa_\text{eff}=\frac{n^\text{ch}}{T}.
\label{eq:k_eff-low-pressure}
\ee
Let us now replace the right-hand side of Eq.~(\ref{eq:derivative-Q}) by Eq.~(\ref{eq:k_eff-low-pressure}), and take the expectation value:
\be
\frac{d\overline{Q_T(P_0+x)}}{dx}=\frac{\overline{n^\text{ch}(x)}}{T}.
\label{eq:expectation-derivative-Q}
\ee
Consistently with our picture, we identify $\overline{n^\text{ch}}$ with $\overline{n_\text{REM}}$ provided in Eq.~(\ref{eq:mean-particle-density-REM}). Integrating Eq.~(\ref{eq:expectation-derivative-Q}) over $x$, taking account of the boundary condition $\overline{Q_T(P_0)}=0$, one easily arrives at Eq.~(\ref{eq:Q-large-T}). The latter is a prediction that we will be able to check numerically.


\subsection{Calculation of selected observables in a numerical implementation of the improved algorithm}

The numerical model is fully defined once the distribution $p(\tau)$ of the elementary thresholds $\tau$ is specified. We expect the asymptotic behavior of the observables we consider to be universal across a broad class of models. The qualitative behavior should not significantly depend on the specific choice of $p(\tau)$, aside from a small number of easily calculable constants, as long as the distribution is sufficiently regular and decreases ``rapidly enough'' for large values of $|\tau|$.

For simplicity, we take the uniform distribution on a set of the contiguous non-negative integers $\{1,2,\cdots,N\}$. The parameter $N$ was set to 20 for the flow calculations presented in Sec.~\ref{subsec:flow}, as these calculations are relatively insensitive to $N$. We increased $N$ to 200 for the observables discussed in Sec.~\ref{subsec:statistics} to minimize discretization artefacts, which these observables are more sensitive to. There would be no problem in principle in taking a continuous distribution $p(\tau)$, but the implementation would be significantly more complicated, technically speaking. The tricky part would be solving Eq.~(\ref{eq:FKPP}) and storing the result for all values of the depth between $0$ and $T$.

The few model-dependent parameters needed as inputs to the analytical formulas are computed in Appendix~\ref{sec:TW}.


\subsubsection{Flow}
\label{subsec:flow}

\begin{figure}
    \centering
    \begin{subfigure}[t]{0.45\textwidth}
    \includegraphics[width=\textwidth]{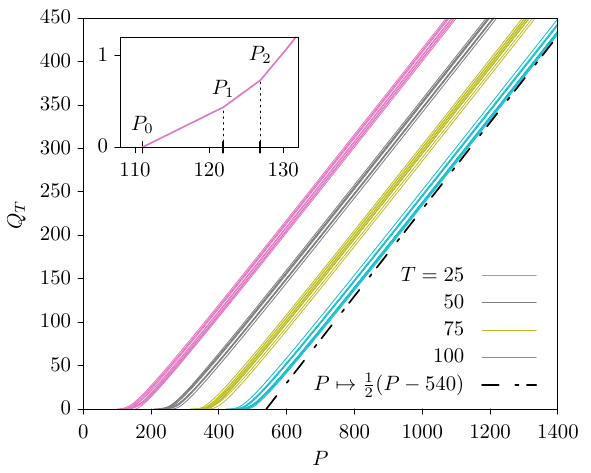}
    \caption{Sets of clusters of 10 realizations of the flow as a function of the pressure $P$ applied at the inlet, for various values of the tree height $T$. The dashed line corresponds to Eq.~(\ref{eq:Q(P)}), with $\kappa_\text{eff}$ set to its asymptotic value at large pressure $\frac12$. (See the main text for a discussion of $P_\text{eff}^*$). {\it Inset:} Zoom in values of $P$ close to $P_0$ for one of the realizations.}
    \label{fig:plot-Q-realizations-a}
    \end{subfigure}
    \hfil
    \begin{subfigure}[t]{0.45\textwidth}
    \includegraphics[width=\textwidth]{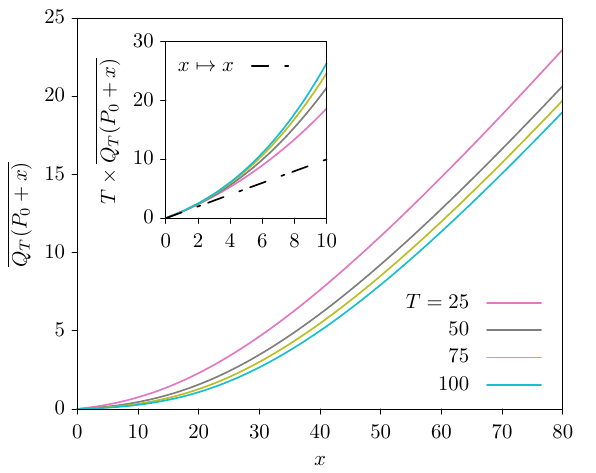}
    \caption{Disorder average of $Q_T(P_0+x)$ over 1000 realizations (to eliminate statistical uncertainties) for different values of $T$, as a function of $x$. {\it Inset:} $\overline{Q_T}$ scaled by $T$ for small values of $x$.}
    \label{fig:plot-Q-realizations-b}
    \end{subfigure}
    \caption{Flow as a function of the applied pressure for different tree heights.}
    \label{fig:plot-Q-realizations}
\end{figure}

As argued in Sec.~\ref{sec:fluid-mechanics}, the flow for a given realization of the network is a piece-wise affine function of the pressure $P$ applied at the inlet, described by the effective Darcy law~(\ref{eq:Q(P)}). In order to visualize $Q_T(P)$, we generated 10 realizations of it for different values of $T$; see Fig.~\ref{fig:plot-Q-realizations-a}. We see that the realizations all exhibit the same global shape, essentially up to a $T$-dependent random shift. The main $T$-dependence must come from the random threshold pressure $P_0$, the expectation value of which roughly grows proportionally to $T$. We can also see clearly that at large $P$, namely for large number of open channels, the slope $\kappa_\text{eff}$ of $Q_T(P)$ for each of the realizations becomes close to the expected asymptotics given in Eq.~(\ref{eq:asymptotics-kappa_eff-P_eff}). It turns out that $P_\text{eff}^*$ instead stays lower than the expected value $T\times\overline{\tau}$, where $\overline{\tau}=10.5$ in our choice for the distribution $p(\tau)$. We shall discuss this fact further below. In Fig.~\ref{fig:plot-Q-realizations-b} is displayed the average of the flow $Q_T(P_0+x)$ over many realizations, as a function of $x$. The curves that correspond to different values of $T$ are very similar at least for large enough $x$, and almost superimpose, now that the realizations have been shifted by the threshold pressure $-P_0$. The inset shows that for large tree heights, the scaled mean flow $T\times\overline{Q_T(P_0+x)}$ may converge to some scaling curve. We also see that the cluster of curves all tend to the function $x\mapsto x$ at the origin $x\to 0$, consistently with the expression~(\ref{eq:Q_T(P)-1-branch}) of $Q_T$ just above the threshold for the opening of the network.

\begin{figure}
    \centering
    \includegraphics[width=0.8\textwidth]{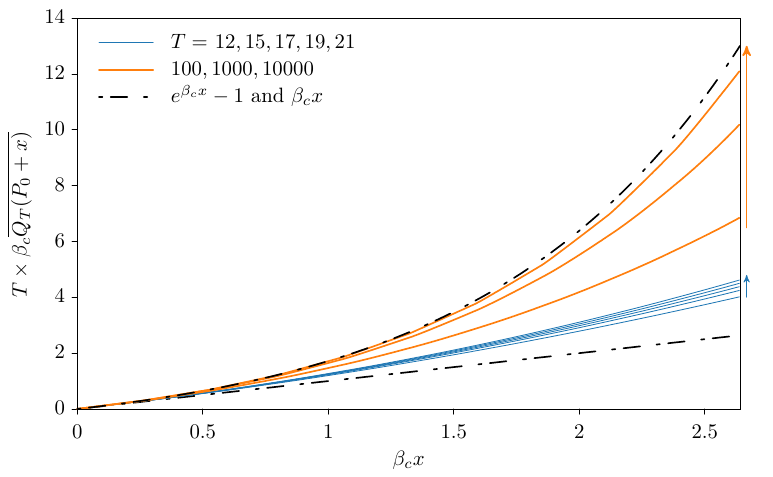}
    \caption{Mean flow for moderate pressure differences. The average is performed over $10^5$ realizations of the network for $T\leq 1000$, and $10^4$ for $T=10000$. In this and some of the following figures, the arrows on the right side of the plots indicate the direction of increasing $T$.}
    \label{fig:plot-Q}
\end{figure}

Let us focus on this mean scaled flow for moderate values of $x$. We perform the calculation with the values of $T$ used in Ref.~\cite{Schimmenti:2023} (i.e. $T\leq 21$), as well as for much larger $T$: for this observable, we were able to generate data for values of $T$ that are about three orders of magnitude larger than in the latter reference. The result of our calculation is shown in Fig.~\ref{fig:plot-Q}. With the new data, we see a clear convergence to the expected asymptotic expression~(\ref{eq:Q-large-T}), which was not possible to see with lower values of $T$: indeed, this convergence turns out to be very slow with $T$.


\paragraph{Effective parameters.}

In order to investigate the properties of the flow in more details, we compute the permeability $\kappa_\text{eff}$ and the threshold pressure $P_\text{eff}^*$ entering the effective Darcy law~(\ref{eq:Q(P)}) as a function of the number of open channels. We then average these quantities over the realizations. 

\begin{figure}
\centering
\begin{subfigure}[t]{0.45\textwidth}
\centering
\includegraphics[width=\textwidth]{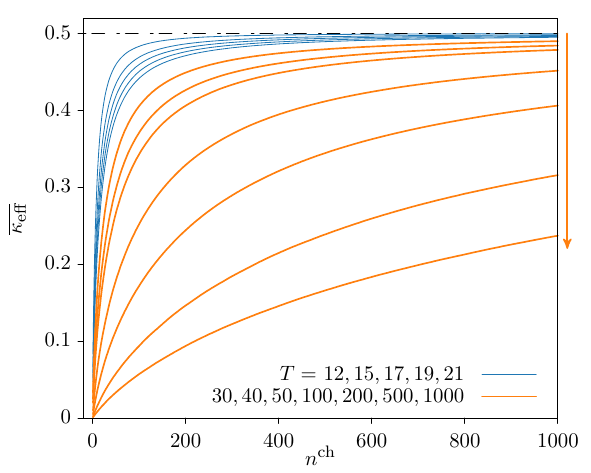}
    \caption{Mean effective permeability as a function of the number of open channels, for different values of the tree height $T$, up to $T=1000$. The expectation values were taken on ensembles of 1000 realizations for $T\leq 500$ and 100 for $T=1000$.}
    \label{fig:plot-keff}
\end{subfigure}
\hfil
\begin{subfigure}[t]{0.45\textwidth}
\centering
\includegraphics[width=\textwidth]{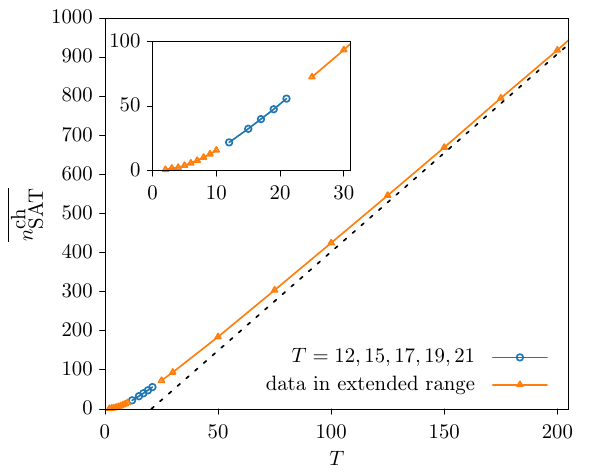}
    \caption{Expectation value of the minimum number of channels $n^\text{ch}_\text{SAT}$ to open such that $\kappa_\text{eff}\geq 0.4$, as a function of the tree height $T$. We display also a straight line to guide the eye. {\it Inset}: Zoom into the lowest values of~$T$.}
    \label{fig:plot-neff}
\end{subfigure}
    \caption{Mean effective permeability as a function of the number of open channels, and mean number of open channels to reach $\kappa_\text{eff}\geq 0.4$.}
    \label{fig:plot-effective-parameters}
\end{figure}

Our numerical data for $\overline{\kappa_\text{eff}}$ is displayed in Fig.~\ref{fig:plot-keff}. We see that the asymptotics~(\ref{eq:asymptotics-kappa_eff-P_eff}) are clearly approached when $T$ assumes moderate values. For larger values of $T$, the expected convergence is slower. For the largest values of $T$ we set ($T=1000$), the limit is hardly approached even when as many as $n^\text{ch}=1000$ channels are open.

In order to quantify the convergence, we compute the expectation value of the minimum number of channels $n^\text{ch}_\text{SAT}$ needed to reach an effective permeability $\kappa_\text{eff}\geq 0.4$; see Fig.~\ref{fig:plot-neff}. We find that the $T$-dependence of this quantity is linear to a good approximation when $T$ becomes large. This clearly confirms the observation in Ref.~\cite{Schimmenti:2023}. The latter was inferred from data generated with the help of the equivalent of the basic algorithm described in Sec.~\ref{sec:naive-algo}, although we now see that these data were very far from any asymptotic regime.

\begin{figure}
    \centering
    \includegraphics[width=0.55\textwidth]{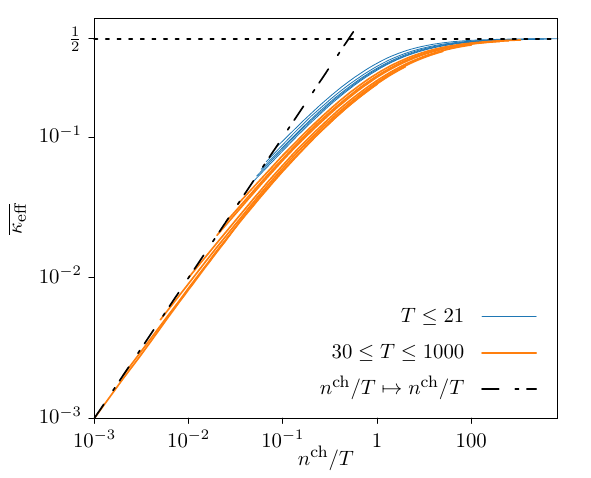}
    \caption{The same data as that shown in Fig.~\ref{fig:plot-keff}, but displayed as a function of the scaling variable $n^\text{ch}/T$. The two expected asymptotes, for $n^\text{ch}\ll T$ and $n^\text{ch}\gg T$, are also drawn.}
    \label{fig:plot-keff-scaling}
\end{figure}

It is also helpful to plot $\overline{\kappa_\text{eff}}$ as a function of the scaled variable $n^\text{ch}/T$; see Fig.~\ref{fig:plot-keff-scaling}. This plot suggests the possibility of convergence to a limiting curve at large $T$. The noticeable dispersion among the cluster of curves indicates a slow convergence. The shapes observed for both low and high numbers of open channels are consistent with the asymptotic form we expect: we approach Eq.~(\ref{eq:k_eff-low-pressure}) in the regime $n^\text{ch}\ll T$, and Eq.~(\ref{eq:asymptotics-kappa_eff-P_eff}) when $n^\text{ch}\gg T$. The transition between these two regimes is confirmed to happen when the number of open channels is on the order of $T$~\cite{Schimmenti:2023}.

\begin{figure}
\centering
\includegraphics[width=.55\textwidth]{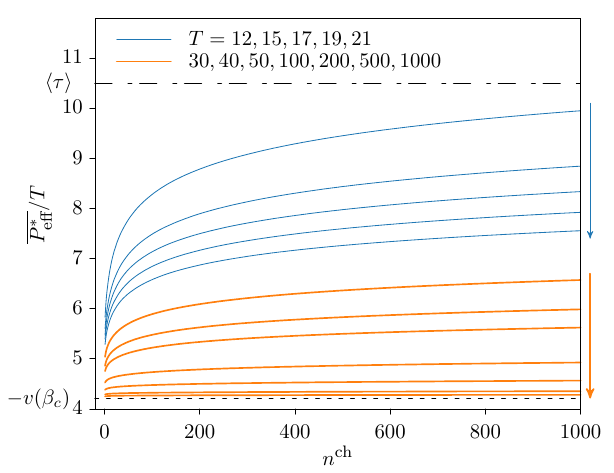}
    \caption{Mean effective threshold pressure scaled by $1/T$ as a function of the number of open channels, for different values of the tree height $T$ up to $T=1000$. We used the same data as in Fig.~\ref{fig:plot-keff}.}
    \label{fig:plot-Peff}
\end{figure}

The mean threshold pressure $\overline{P_\text{eff}^*}$ is displayed in Fig.~\ref{fig:plot-Peff}, as a function of the number of open channels $n^\text{ch}$ and for different values of $T$. When $n^\text{ch}$ is large, we expect that $\overline{P_\text{eff}^*}/T$ tends to $\overline{\tau}$, as observed in Fig.~\ref{fig:plot-Peff} for fixed (low) values of $T$. However, when $T$ is taken large while keeping $n^\text{ch}$ fixed, this quantity seems to converge instead to $-v(\beta_c)$, namely to ${\overline{P_0}}/{T}$ (up to terms on the order of ${\ln T}/{T}$; see Eq.~(\ref{eq:keff-Peff-nch=1}) and Eq.~(\ref{eq:position-TW})). This fact might seem puzzling at first sight, but is actually easy to understand. Indeed, the total number of channels grows like $2^T$, hence for large values of $T$, the $n^\text{ch}\sim T$ first channels to open belong to the ``low-energy'' part of the ``energy spectrum'', in the context of the DP interpretation. Consequently, as $T$ increases while $n^\text{ch}$ remains fixed, the system is effectively constrained to states with the lowest energy, on the order of $\overline{\varepsilon_{0\,T}^{({\cal P}_0)}}=\overline{P_0}$.

\subsubsection{Statistics of the number of open channels}
\label{subsec:statistics}

So far, we have focused on the flow, the most natural observable for this problem. Now we focus on $n^{\text{ch}}(x)$, namely the number of open channels at the pressure $P= P_0+x$. Note that here we fix the value of $x$ for all disorder realization while the threshold $P_0$ and thus the pressure $P$ are realization dependent. In particular we are interested in its average $\overline{n^{\text{ch}}(x)}$ and its distribution. 

\begin{figure}
\centering
    \begin{subfigure}[t]{0.45\textwidth}
    \centering
    \includegraphics[width=\textwidth]{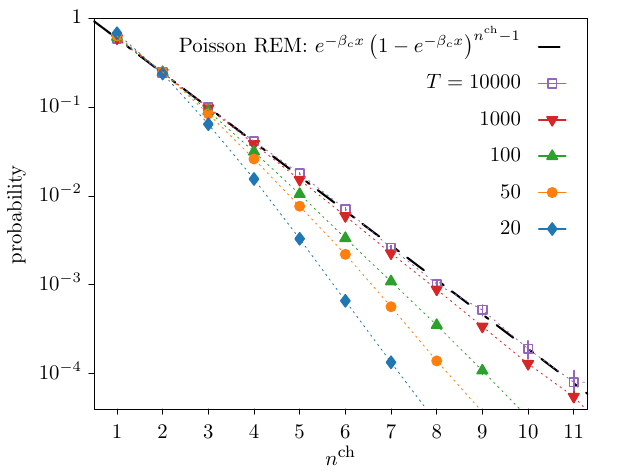}
    \caption{Distribution of $n^{\text{ch}}$ for $x=20$ (i.e. $\beta_c x=0.5262\cdots$) for different sizes $T$, compared to the Poisson-REM prediction~(\ref{eq:Poisson-REM-body}). We generated $10^6$ realizations for $T\leq 1000$ and $10^5$ realizations for $T=10000$.}
    \label{fig:plot-nch-histo}
    \end{subfigure}
    \hfil
    \begin{subfigure}[t]{0.45\textwidth}
    \centering
    \includegraphics[width=\textwidth]{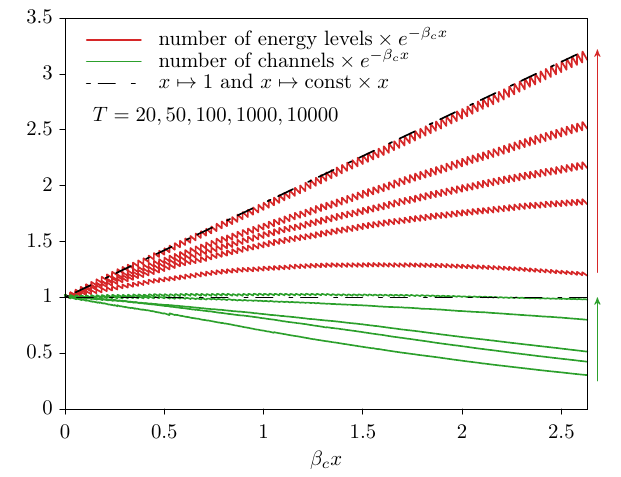}
    \caption{Mean number of open channels, $\overline {n^{\text{ch}}(x)}$, and mean number of energy levels such that $\varepsilon-\varepsilon_0\leq x$, as a function of $\beta_c x$. Both quantities are divided by $\exp(\beta_c x)$. In each set, the different lines correspond to different values of $T$. The arrows indicate the direction of increasing $T$. Averages are taken over ensembles of $10^4$ realizations.}
    \label{fig:plot-nch-mean}
    \end{subfigure}
    \caption{Statistics of the number of open channels in the inlet pressure interval $[P_0,P_0+x]$ or of energy levels in the energy interval $[\varepsilon_0,\varepsilon_0+x]$.}
    \label{fig:plot-nch}
\end{figure}

In Fig.~\ref{fig:plot-nch-histo} is displayed the histogram of $n^{\text{ch}}(x)$ for  $x=20$. We see that the larger $T$, the closer one gets from what would be expected for the Random Energy Model (REM), namely
\be
P_{n^\text{ch}}(x)=e^{-\beta_c x}\left(1-e^{-\beta_c x}\right)^{n^{\text{ch}}-1}\,,
\label{eq:Poisson-REM-body}
\ee
consistently with the simplified picture (see Appendix~\ref{sec:Poisson-REM}). We have set $N=200$, instead of $N=20$ used for the flow calculation: in this way, the degeneracies  of the energy levels due to finite $N$ are strongly suppressed and we are able to recover the asymptotic large $T$  behavior given in  Eqs.~(\ref{eq:Poisson-REM}),(\ref{eq:Poisson-REM-body}).

Figure~\ref{fig:plot-nch-mean} shows the mean, $\overline{n^\text{ch}(x)}$, divided by what the expected  leading behavior, namely $e^{\beta_c x}$. We see that the larger $T$,  the curves get constant  and closer to 1. Again, the convergence is quite slow. We deemed it useful to compare the mean number of open channels to the mean number of configurations of the DP in the energy interval $x$ above the ground state. We see that the latter also converges to the expected asymptotics $\propto x \,e^{\beta_c x}$.

Note that the small  oscillations seen in the curves are an effect of finite  $N$ and thus of the  fact that the energy levels are discrete. Their  amplitude increases for a smaller $N$. However, we have verified that also for $N=20$ used for the flow the trend of the curves is the same.

\section{Conclusion}
\label{sec:conclusion}

We have designed and implemented a novel algorithm that exactly generates the open channels of a tree-like network of throats at a specified pressure. This method is effectively a spinal decomposition of the tree, with the primary spine corresponding to the first open channel. Leveraging this algorithm, we were able to analyze networks of unprecedented depth, achieving a two to three orders of magnitude increase in tree height $T$ compared to previous studies, all while using a standard laptop. Given the algorithm's linear complexity with respect to $T$, even greater heights are within reach.

The results obtained from our algorithm confirm with remarkable precision the conjectures proposed in Ref.~\cite{Schimmenti:2023}. Specifically, we have successfully verified the analytical expression for the mean flow [Eq.(\ref{eq:Q-large-T})], demonstrating the feasibility of deriving an explicit expression for the non-linear Darcy law of a yield stress fluid within this network. Moreover, our findings strongly support the physical picture argued in Ref.~\cite{Schimmenti:2023}:
\begin{itemize}
    \item  The first channels that open are those with low opening thresholds and minimal overlap. As a consequence, the number of open channels, \(n^\text{ch}(x)\), identifies to the number of low-energy levels of the Random Energy Model (REM).
    
    \item The effective permeability, \(\kappa_{\text{eff}}\), initially increases linearly with the number of open channels. Upon the activation of approximately \(T\) channels, \(\kappa_{\text{eff}}\) saturates to the Newtonian permeability of the medium. It is noteworthy that \(T\) channels represent a very small fraction compared to the total possible channels in the network, \(2^{T-1}\).
    
    \item  The effective offset, \(P_{\text{eff}}^*\), gradually evolves from the ground state energy of the associated directed polymer to the mean energy of a non-optimized polymer.
\end{itemize}

These data and our algorithm open up the prospect of studying sub-asymptotic corrections, and to elaborate a more refined picture of the corresponding physical effects, beyond the simplified REM model for the tree-like network. These results provide a robust validation of the theoretical framework and offer significant insights into the behavior of yield stress fluids in complex networks.

A longer-term and more ambitious project would be to understand if the behavior of the effective permeability and effective  offset is peculiar to tree-like networks or if it generalizes to networks of finite dimension~\cite{Liu:2019}.


\section*{Acknowledgments}

S.M. thanks Pascal Maillard, Bastien Mallein, Michel Pain for their interest and for related discussions. A.R. thanks Chen Liu, A. De Luca, V. Schimmenti and L. Talon for the useful discussions.  This project was initiated in the framework of the ``GDR Branchement''. It has received financial support from the CNRS through the MITI interdisciplinary programs, and from the Agence Nationale de la Recherche (ANR), grant ANR-23-CE30-0031-04 (DISCREEP).


\appendix

\section{Traveling wave solutions to the non-linear equation and energy levels of the directed polymer near the ground state}
\label{sec:TW}

From Eq.~(\ref{eq:FKPP}) one derives the probability distribution of the ground state of the directed polymer. Let us discuss the solutions to this equation.

It is well-known that for a wide class of initial conditions, including those relevant to this paper, the large-time solution of the non-linear equation~(\ref{eq:FKPP}) forms a universal traveling wave (see e.g. Ref.~\cite{VanSaarloos:2003}, and Ref.~\cite{Bramson:1983} for the mathematical proof in the case of the FKPP equation). The latter refers to a uniformly-moving front that monotonously connects $u_s=0$ (for $\varepsilon'\to-\infty$) to $u_s=1$ (for $\varepsilon'\to+\infty$). The standard way of determining the properties of the traveling wave is to solve the equation obtained from the linearization of Eq.~(\ref{eq:FKPP}) in the region where $u_s$ is small, where only the first term in the right-hand side is relevant. A suitable Ansatz is
\be
u_s(\varepsilon')\propto e^{\beta[\varepsilon'+v(\beta)s]},
\label{eq:Ansatz-TW}
\ee
where $\beta$ is a positive parameter. Requiring that this form be a solution of the linearized equation~(\ref{eq:FKPP}) determines $v(\beta)$:
\be
v(\beta)=\frac{1}{\beta}\ln\left(2\sum_\tau p(\tau)e^{-\beta\tau}\right).
\label{eq:v}
\ee
(See e.g. \cite{Derrida:2016} for a discussion of a very similar model). For the initial condition in Eq.~(\ref{eq:FKPP}), the shape of the asymptotic traveling wave in the region where $u_s(\varepsilon')\ll 1$ is known to be of the form~(\ref{eq:Ansatz-TW}) with $\beta=\beta_c$, where $\beta_c$ minimizes $v(\beta)$. The velocity of the wave is then $-v(\beta_c)$.

Corrections for large but finite network heights $s$ are also known. The front position reads~\cite{Bramson:1983}
\be
m_s=-v(\beta_c)s+\frac{3}{2\beta_c}\ln s+\text{const}+o(1)\,.
\label{eq:position-TW}
\ee
This quantity also represents the expectation value (or, up to a constant, the median value) of the ground-state energy $\varepsilon_0^\text{bin}$ of a directed polymer on a binary tree of height $s$. For $\varepsilon'<m_s$ and in the limit $m_s-\varepsilon'\gg 1$, the traveling wave has the following shape:
\be
u_s(\varepsilon')\simeq\text{const}\times\left(m_s-\varepsilon'+{\text{const}}'\right) e^{\beta_c(\varepsilon'-m_s)}\times e^{-{(\varepsilon'-m_s)^2}/[2\beta_c v''(\beta_c)s]}\,.
\ee
In the region $m_s-\varepsilon'\ll\sqrt{\beta_c v''(\beta_c)s}$, the dominant $\varepsilon'$-dependence is indeed the exponential form~(\ref{eq:Ansatz-TW}), with $\beta=\beta_c$, corrected by a linear factor. For $m_s-\varepsilon'\gtrsim\sqrt{\beta_c v''(\beta_c)s}$, the Gaussian factor drives $u_s(P')$ rapidly to very small values.

A derivation of these results can be found e.g. in Ref.~\cite{Brunet:1997}. It is not difficult to recover them heuristically: it is sufficient to recognize that, at this level of accuracy and for the particular quantities considered, the non-linear evolution equation can be replaced by its linearization, supplemented with a moving absorptive boundary. We refer the reader to the established literature on this topic, e.g. Ref.~\cite{VanSaarloos:2003} and references therein.


\paragraph{Mean density of energy levels.}

Analyzing further the relationship between the FKPP equation and branching Brownian motion, or between the corresponding finite-difference equation and its associated branching random walk respectively, one may prove that the full information about the distribution of the energy levels near the ground state can be deduced from the solution to Eq.~(\ref{eq:FKPP}), if one takes appropriate initial conditions~\cite{Brunet:2011,Aidekon:2013}. Of particular interest for the present work is the expectation value $\overline{n_\text{BRW}(x)}$ of the number of energy levels at a distance $x$ above the ground-state energy $\varepsilon_0^\text{bin}$~\cite{Brunet:2011,Mueller:2019ror}:
\be
\overline{n_\text{BRW}(x)}\simeq\text{const}\times x\, e^{\beta_c x}\,.
\label{eq:mean-particle-density-FKPP}
\ee
This formula holds for (formally) infinite-$s$, and $x\gg 1/\beta_c$.


\paragraph{Choice of $p(\tau)$ in our numerical calculations.}

In our particular implementation, we chose $p(\tau)$ uniform on the first $N$ non-zero integers, namely
\be
p(\tau)=\begin{cases} {1}/{N} &  \text{for $\tau\in\{1,2\cdots N\}$},\\
0 &\text{else}.
\end{cases}
\ee
Concretely, for the two values of $N$ we have used for our numerical simulations, the relevant parameters are given in Tab.~\ref{tab:pars}.

\begin{table}[ht]
    \centering
    \begin{tabular}{rrrr}
        \toprule
        $N$ & \multicolumn{1}{c}{$N\beta_c$} & \multicolumn{1}{c}{$-v(\beta_c)/N$} & \multicolumn{1}{c}{$\beta_c v''(\beta_c)/N^2$} \\
        \midrule
        $20$  & $5.27993\cdots$ & $0.210376\cdots$ & $0.0305183\cdots$ \\
        $200$ & $5.26225\cdots$ & $0.187333\cdots$ & $0.0308726\cdots$ \\
        $\infty$ & $5.26207\cdots$ & $0.184827\cdots$ & $0.0308761\cdots$ \\
        \bottomrule
    \end{tabular}
    \caption{Values of the relevant parameters used in the numerical calculations (first two lines). We applied a scaling $\tau\mapsto \tilde\tau=\tau/N$ such that for finite $N$, $\tilde\tau \mapsto N\,p(N\tilde\tau)$ is a discretization of the square function $\tilde\tau\mapsto {\mathbbm 1}_{\{0\leq \tilde\tau\leq 1\}}$. The last line, $N\to\infty$, represents the continuous limit, which we give so that one can appreciate the distance with the finite $N$.}
    \label{tab:pars}
\end{table}


\section{Energy levels in a Random Energy Model}
\label{sec:Poisson-REM}

Let us assume that our configuration space consists of $2^{T-1}$ independent branches, and consider a branch labeled ${\cal P}$. It is characterized by the random energy $\varepsilon_{0\,T}^{({\cal P})}$. The generating function $G(\gamma)$ of the probability $\P(\varepsilon_{0\,T}^{({\cal P})}=\varepsilon)$ that this energy equals some given $\varepsilon$ reads
\be
G(\gamma)\equiv\sum_\varepsilon e^{-\gamma\varepsilon}\,\P(\varepsilon_{0\,T}^{({\cal P})}=\varepsilon)
=\sum_\varepsilon e^{-\gamma\varepsilon}\left[\sum_{\tau'_{0\,1},\cdots,\tau'_{T-1\,T}\in {\cal P}}\left(\prod_{j=0}^{T-1} p(\tau'_{j\,j+1})\right)\delta_{\varepsilon,\sum_{j=0}^{T-1}\tau'_{j\,j+1}}\right],
\ee
and is obviously the same for all branches. An easy calculation yields
\be
G(\gamma)=\left(\sum_{\tau'}p(\tau')e^{-\gamma\tau'}\right)^T.
\ee
The sought probability is then obtained from an appropriate integration:
\be
\P(\varepsilon_{0\,T}^{({\cal P})}=\varepsilon)=\int_{\cal C} \frac{d\gamma}{2i\pi}\,e^{\gamma\varepsilon}G(\gamma)=\int_{\cal C} \frac{d\gamma}{2i\pi}\,e^{\gamma\varepsilon}\left(\sum_{\tau'}p(\tau')e^{-\gamma\tau'}\right)^T\,,
\label{eq:proba-energy-REM}
\ee
where the integration domain can be, for example, the segment ${\cal C}\equiv[\gamma_0-i\pi,\gamma_0+i\pi]$ in the complex plane, for any real number $\gamma_0$. In the limit of very large $T$, the integral may be evaluated by a steepest-descent method. We introduce
\be
\chi(\gamma)\equiv \ln\left(2\sum_{\tau'}p(\tau')e^{-\gamma\tau'}\right)
\label{eq:chi}
\ee
in terms of which the probability~(\ref{eq:proba-energy-REM}) can be conveniently re-expressed:
\be
\P(\varepsilon_{0\,T}^{({\cal P})}=\varepsilon)=\int_{\cal C} \frac{d\gamma}{2i\pi}\,e^{\gamma\varepsilon+\left[\chi(\gamma)-\ln 2\right]T}.
\ee
We denote by $\gamma_c$ the solution to the saddle-point equation
\be
\varepsilon+\chi'(\gamma_c)T=0\,.
\label{eq:saddle-point}
\ee
We then set $\gamma_0=\gamma_c$, and observe that the integration domain ${\cal C}$ may be extended to an infinite line parallel to the imaginary axis in the complex plane: the contribution to the integral of the added pieces of contour would be small for large $T$. The steepest-descent method evaluation of $\P(\varepsilon_{0\,T}^{({\cal P})}=\varepsilon)$ yields
\be
\P(\varepsilon_{0\,T}^{({\cal P})}=\varepsilon)\underset{T\gg 1}{\sim} e^{\gamma_c\varepsilon+\left[\chi(\gamma_c)-\ln 2\right]T}\,.
\label{eq:p(eps)-saddle-point}
\ee
From Eq.~(\ref{eq:p(eps)-saddle-point}), we see that $\P(\varepsilon_{0\,T}^{({\cal P})}=\varepsilon)$ decreases exponentially as $\varepsilon$ decreases. Thus the typical ground state energy $\varepsilon_0$ is found by requiring that the probability that there is no state of that energy among the $2^{T-1}$ independent branches is of order unity (for example $\frac12$; the precise value is actually irrelevant). This probability reads
\be
\left[1-\P(\varepsilon_{0\,T}^{({\cal P})}=\varepsilon_0)\right]^{2^{T-1}}\,.
\ee
Asking for this quantity to be of order unity for $T$ large is equivalent to requiring that $2^T \P(\varepsilon_{0\,T}^{({\cal P})}=\varepsilon_0)={\cal O}(1)$. Taking the logarithm and using the saddle-point evaluation of $\P(\varepsilon_{0\,T}^{({\cal P})}=\varepsilon_0)$ in Eq.~(\ref{eq:p(eps)-saddle-point}), neglecting terms that are small compared to $T$, we find
\be
\gamma_c\varepsilon_0+\chi(\gamma_c)T=0\,.
\label{eq:fixed-p}
\ee
Combining Eq.~(\ref{eq:fixed-p}) and Eq.~(\ref{eq:saddle-point}), up to the replacement $\varepsilon\to\varepsilon_0$ in the latter, we see that necessarily 
\be
\gamma_c\chi'(\gamma_c)=\chi(\gamma_c).
\label{eq:gamma_c}
\ee 
Comparing $\chi$ defined in Eq.~(\ref{eq:chi}) and $v$ introduced in Appendix~\ref{sec:TW} [Eq.~(\ref{eq:v})], we observe that $\chi(\gamma)\equiv\gamma v(\gamma)$. Hence, Eq.~(\ref{eq:gamma_c}) is equivalent to $v'(\gamma_c)=0$, which implies that $\gamma_c=\beta_c$. The mean density of levels of energy $\varepsilon$ near the typical ground state energy $\overline{\varepsilon_0}\simeq -v(\beta_c)T$ then reads
\be
2^{T-1}\P(\varepsilon_{0\,T}^{({\cal P})}=\varepsilon)\simeq e^{\beta_c(\varepsilon-\overline{\varepsilon_0})}.
\label{eq:mean-number-levels-REM}
\ee

If $T$ is large enough, we can consider that the distribution of the energy levels is Poissonian~\cite{Derrida:2015}, the rate parameter of the Poisson law being the exponential in the right-hand side of Eq.~(\ref{eq:mean-number-levels-REM}).

One may compute the probability $P_n(x)$ of having $n$ energy levels in the interval of size $x$ above the ground state energy, namely in $[\varepsilon_0,\varepsilon_0+x]$. When $\beta_c$ is small, which is verified for large $N$, then $x$ can be considered a continuous variable, since the rate parameter of the Poisson process varies slowly with $x$. The latter becomes the intensity of a Poisson point-like process on the line. In this continuous limit, we get a simple expression for $P_n(x)$:
\be
P_n(x)=e^{-\beta_c x}\left(1-e^{-\beta_c x}\right)^{n-1}\,.
\label{eq:Poisson-REM}
\ee
The mean number of levels which follows from this probability reads
\be
\overline{n_\text{REM}(x)}=e^{\beta_c x}.
\label{eq:mean-particle-density-REM}
\ee
Comparing this formula to the equivalent one for branching random walks~(\ref{eq:mean-particle-density-FKPP}), we see that they differ essentially by a linear factor $x$.



\end{document}